\RequirePackage{lineno}
\documentclass[twocolumn]{aastex61}
\pdfoutput=1
\usepackage{amssymb}
\usepackage{gensymb}
\usepackage{graphicx}
\usepackage{amsmath}
\usepackage{wasysym}
\usepackage{verbatim}
\usepackage{rotating}
\usepackage{subfigure}

\submitjournal{ApJ}

\begin{document}

\title{Dark Matter Limits from Dwarf Spheroidal Galaxies with the HAWC Gamma-Ray Observatory}

\correspondingauthor{Tolga Yapici}
\email{tyapici@pa.msu.edu}

\author{A.~Albert}
\affil{Physics Division, Los Alamos National Laboratory, Los Alamos, NM, USA }
\author{R.~Alfaro}
\affil{Instituto de F\'{i}sica, Universidad Nacional Aut\'{o}noma de Mexico, Mexico City, Mexico }
\author{C.~Alvarez}
\affil{Universidad Aut\'{o}noma de Chiapas, Tuxtla Guti\'{e}rrez, Chiapas, Mexico}
\author{J.D.~\'{A}lvarez}
\affil{Universidad Michoacana de San Nicol\'{a}s de Hidalgo, Morelia, Mexico }
\author{R.~Arceo}
\affil{Universidad Aut\'{o}noma de Chiapas, Tuxtla Guti\'{e}rrez, Chiapas, Mexico}
\author{J.C.~Arteaga-Vel\'{a}zquez}
\affil{Universidad Michoacana de San Nicol\'{a}s de Hidalgo, Morelia, Mexico }
\author{D.~Avila Rojas}
\affil{Instituto de F\'{i}sica, Universidad Nacional Aut\'{o}noma de M\'{e}xico, Mexico City, Mexico }
\author{H.A.~Ayala Solares}
\affil{Department of Physics, Michigan Technological University, Houghton, MI, USA }
\author{N.~Bautista-Elivar}
\affil{Universidad Politecnica de Pachuca, Pachuca, Hidalgo, Mexico }
\author{A.~Becerril}
\affil{Instituto de F\'{i}sica, Universidad Nacional Aut\'{o}noma de M\'{e}xico, Mexico City, Mexico }
\author{E.~Belmont-Moreno}
\affil{Instituto de F\'{i}sica, Universidad Nacional Aut\'{o}noma de M\'{e}xico, Mexico City, Mexico }
\author{S.Y.~BenZvi}
\affil{Department of Physics \& Astronomy, University of Rochester, Rochester, NY , USA }
\author{A.~Bernal}
\affil{Instituto de Astronom\'{i}a, Universidad Nacional Aut\'{o}noma de M\'{e}xico, Mexico City, Mexico }
\author{J.~Braun}
\affil{Department of Physics, University of Wisconsin-Madison, Madison, WI, USA }
\author{K.S.~Caballero-Mora}
\affil{Universidad Aut\'{o}noma de Chiapas, Tuxtla Guti\'{e}rrez, Chiapas, Mexico}
\author{T.~Capistr\'{a}n}
\affil{Instituto Nacional de Astrof\'{i}sica, \'{O}ptica y Electr\'{o}nica, Tonantzintla, Puebla, Mexico }
\author{A.~Carrami\~{n}ana}
\affil{Instituto Nacional de Astrof\'{i}sica, \'{O}ptica y Electr\'{o}nica, Tonantzintla, Puebla, Mexico }
\author{M.~Castillo}
\affil{Universidad Michoacana de San Nicol\'{a}s de Hidalgo, Morelia, Mexico }
\author{U.~Cotti}
\affil{Universidad Michoacana de San Nicol\'{a}s de Hidalgo, Morelia, Mexico }
\author{C.~De Le\'{o}n}
\affil{Facultad de Ciencias F\'{i}sico Matem\'{a}ticas, Benem\'{e}rita Universidad Aut\'{o}noma de Puebla, Puebla, Mexico }
\author{E.~De la Fuente}
\affil{Departamento de F\'{i}sica, Centro Universitario de Ciencias Exactas e Ingenier\'{i}as, Universidad de Guadalajara, Guadalajara, Mexico }
\author{R.~Diaz Hernandez}
\affil{Instituto Nacional de Astrof\'{i}sica, \'{O}ptica y Electr\'{o}nica, Tonantzintla, Puebla, Mexico }
\author{B.L.~Dingus}
\affil{Physics Division, Los Alamos National Laboratory, Los Alamos, NM, USA }
\author{M.A.~DuVernois}
\affil{Department of Physics, University of Wisconsin-Madison, Madison, WI, USA }
\author{J.C.~D\'{i}az-V\'{e}lez}
\affil{Departamento de F\'{i}sica, Centro Universitario de Ciencias Exactas e Ingenier\'{i}as, Universidad de Guadalajara, Guadalajara, Mexico }
\author{R.W.~Ellsworth}
\affil{School of Physics, Astronomy, and Computational Sciences, George Mason University, Fairfax, VA, USA }
\author{D.W.~Fiorino}
\affil{Department of Physics, University of Maryland, College Park, MD, USA }
\author{N.~Fraija}
\affil{Instituto de Astronom\'{i}a, Universidad Nacional Aut\'{o}noma de M\'{e}xico, Mexico City, Mexico }
\author{J.A.~Garc\'{i}a-Gonz\'{a}lez}
\affil{Instituto de F\'{i}sica, Universidad Nacional Aut\'{o}noma de M\'{e}xico, Mexico City, Mexico }
\author{F.~Garfias}
\affil{Instituto de Astronom\'{i}a, Universidad Nacional Aut\'{o}noma de M\'{e}xico, Mexico City, Mexico }
\author{M.M.~Gonz\'{a}lez}
\affil{Instituto de Astronom\'{i}a, Universidad Nacional Aut\'{o}noma de M\'{e}xico, Mexico City, Mexico }
\author{J.A.~Goodman}
\affil{Department of Physics, University of Maryland, College Park, MD, USA }
\author{J.P.~Harding}
\affil{Physics Division, Los Alamos National Laboratory, Los Alamos, NM, USA }
\author{S.~Hernandez}
\affil{Instituto de F\'{i}sica, Universidad Nacional Aut\'{o}noma de M\'{e}xico, Mexico City, Mexico }
\author{A.~Hernandez-Almada}
\affil{Instituto de F\'{i}sica, Universidad Nacional Aut\'{o}noma de M\'{e}xico, Mexico City, Mexico }
\author{A.~Iriarte}
\affil{Instituto de Astronom\'{i}a, Universidad Nacional Aut\'{o}noma de M\'{e}xico, Mexico City, Mexico }
\author{V.~Joshi}
\affil{Max-Planck Institute for Nuclear Physics, Heidelberg, Germany}
\author{S.~Kaufmann}
\affil{Universidad Aut\'{o}noma de Chiapas, Tuxtla Guti\'{e}rrez, Chiapas, Mexico}
\author{D.~Kieda}
\affil{Department of Physics and Astronomy, University of Utah, Salt Lake City, UT, USA }
\author{R.J.~Lauer}
\affil{Department of Physics and Astronomy, University of New Mexico, Albuquerque, NM, USA }
\author{D.~Lennarz}
\affil{School of Physics and Center for Relativistic Astrophysics - Georgia Institute of Technology, Atlanta, GA, USA}
\author{H.~Le\'{o}n Vargas}
\affil{Instituto de F\'{i}sica, Universidad Nacional Aut\'{o}noma de M\'{e}xico, Mexico City, Mexico }
\author{J.T.~Linnemann}
\affil{Department of Physics and Astronomy, Michigan State University, East Lansing, MI, USA }
\author{A.L.~Longinotti}
\affil{Instituto Nacional de Astrof\'{i}sica, \'{O}ptica y Electr\'{o}nica, Tonantzintla, Puebla, Mexico }
\author{G.~Luis Raya}
\affil{Universidad Politecnica de Pachuca, Pachuca, Hidalgo, Mexico }
\author{R.~L\'{o}pez-Coto}
\affil{Max-Planck Institute for Nuclear Physics, Heidelberg, Germany}
\author{K.~Malone}
\affil{Department of Physics, Pennsylvania State University, University Park, PA, USA }
\author{S.S.~Marinelli}
\affil{Department of Physics and Astronomy, Michigan State University, East Lansing, MI, USA }
\author{I.~Martinez-Castellanos}
\affil{Department of Physics, University of Maryland, College Park, MD, USA }
\author{J.~Mart\'{i}nez-Castro}
\affil{Centro de Investigaci\'on en Computaci\'on, Instituto Polit\'ecnico Nacional, Mexico City, Meexico.}
\author{H.~Mart\'{i}nez-Huerta}
\affil{Physics Department, Centro de Investigacion y de Estudios Avanzados del IPN, Mexico City, Mexico }
\author{J.A.~Matthews}
\affil{Department of Physics and Astronomy, University of New Mexico, Albuquerque, NM, USA }
\author{P.~Miranda-Romagnoli}
\affil{Universidad Aut\'{o}noma del Estado de Hidalgo, Pachuca, Mexico }
\author{E.~Moreno}
\affil{Facultad de Ciencias F\'{i}sico Matem\'{a}ticas, Benem\'{e}rita Universidad Aut\'{o}noma de Puebla, Puebla, Mexico }
\author{L.~Nellen}
\affil{Instituto de Ciencias Nucleares, Universidad Nacional Aut\'{o}noma de Mexico, Mexico City, Mexico }
\author{M.~Newbold}
\affil{Department of Physics and Astronomy, University of Utah, Salt Lake City, UT, USA }
\author{M.U.~Nisa}
\affil{Department of Physics \& Astronomy, University of Rochester, Rochester, NY , USA }
\author{R.~Noriega-Papaqui}
\affil{Universidad Aut\'{o}noma del Estado de Hidalgo, Pachuca, Mexico }
\author{R.~Pelayo}
\affil{Centro de Investigaci\'on en Computaci\'on, Instituto Polit\'ecnico Nacional, Mexico City, M\'exico.}
\author{J.~Pretz}
\affil{Department of Physics, Pennsylvania State University, University Park, PA, USA }
\author{E.G.~P\'{e}rez-P\'{e}rez}
\affil{Universidad Politecnica de Pachuca, Pachuca, Hidalgo, Mexico }
\author{Z.~Ren}
\affil{Department of Physics and Astronomy, University of New Mexico, Albuquerque, NM, USA }
\author{C.D.~Rho}
\affil{Department of Physics \& Astronomy, University of Rochester, Rochester, NY , USA }
\author{C.~Rivi\`{e}re}
\affil{Department of Physics, University of Maryland, College Park, MD, USA }
\author{D.~Rosa-Gonz\'{a}lez}
\affil{Instituto Nacional de Astrof\'{i}sica, \'{O}ptica y Electr\'{o}nica, Tonantzintla, Puebla, Mexico }
\author{E.~Ruiz-Velasco}
\affil{Instituto de F\'{i}sica, Universidad Nacional Aut\'{o}noma de M\'{e}xico, Mexico City, Mexico }
\author{F.~Salesa Greus}
\affil{Instytut Fizyki Jadrowej im Henryka Niewodniczanskiego Polskiej Akademii Nauk, Krakow, Poland }
\author{A.~Sandoval}
\affil{Instituto de F\'{i}sica, Universidad Nacional Aut\'{o}noma de M\'{e}xico, Mexico City, Mexico }
\author{M.~Schneider}
\affil{Santa Cruz Institute for Particle Physics, University of California, Santa Cruz, Santa Cruz, CA, USA }
\author{H.~Schoorlemmer}
\affil{Max-Planck Institute for Nuclear Physics, Heidelberg, Germany}
\author{G.~Sinnis}
\affil{Physics Division, Los Alamos National Laboratory, Los Alamos, NM, USA }
\author{A.J.~Smith}
\affil{Department of Physics, University of Maryland, College Park, MD, USA }
\author{R.W.~Springer}
\affil{Department of Physics and Astronomy, University of Utah, Salt Lake City, UT, USA }
\author{P.~Surajbali}
\affil{Max-Planck Institute for Nuclear Physics, Heidelberg, Germany}
\author{I.~Taboada}
\affil{School of Physics and Center for Relativistic Astrophysics - Georgia Institute of Technology, Atlanta, GA, USA}
\author{O.~Tibolla}
\affil{Universidad Aut\'{o}noma de Chiapas, Tuxtla Guti\'{e}rrez, Chiapas, Mexico}
\author{K.~Tollefson}
\affil{Department of Physics and Astronomy, Michigan State University, East Lansing, MI, USA }
\author{I.~Torres}
\affil{Instituto Nacional de Astrof\'{i}sica, \'{O}ptica y Electr\'{o}nica, Tonantzintla, Puebla, Mexico }
\author{T.~Weisgarber}
\affil{Department of Physics, University of Wisconsin-Madison, Madison, WI, USA }
\author{T.~Yapici}
\affil{Department of Physics and Astronomy, Michigan State University, East Lansing, MI, USA }
\author{H.~Zhou}
\affil{Physics Division, Los Alamos National Laboratory, Los Alamos, NM, USA }

\begin{abstract}
The High Altitude Water Cherenkov (HAWC) gamma-ray observatory is a wide field of view observatory sensitive to $500~\rm GeV \-- 100~\rm TeV$ gamma rays and cosmic rays. It can also perform diverse indirect searches for dark matter (DM) annihilation and decay. Among the most promising targets for the indirect detection of dark matter are dwarf spheroidal galaxies. These objects are expected to have few astrophysical sources of gamma rays but high dark matter content, making them ideal candidates for an indirect dark matter detection with gamma rays. Here we present individual limits on the annihilation cross section and decay lifetime for 15 dwarf spheroidal galaxies within the HAWC field-of-view, as well as their combined limit. These are the first limits on the annihilation cross section and decay lifetime using data collected with HAWC.
\end{abstract}

\keywords{astroparticle physics, galaxies: dwarf, dark matter, gamma rays: general}


%
\section{Introduction}
While the evidence for dark matter is ample, there remains the question of its composition. There are numerous dark matter candidates, categorized into non-baryonic and baryonic dark matter. Among the possibilities of non-baryonic dark matter candidates, Weakly Interacting Massive Particles (WIMPs) are one of the leading hypothetical particle physics candidates for cold dark matter. A WIMP is a dark matter particle that has fallen out of thermal equilibrium with the hot dense plasma during the beginning of the Universe and interacts with known standard model particles via a force similar in strength to the weak force \citep{SUSYdm}. In dense dark matter regions, WIMPs can annihilate into Standard Model particles. The products of the annihilation can produce photons via pion decay, radiative processes by charged leptons, or direct production of gamma rays through loop processes.

The expected dark matter annihilation cross-section depends on the exact model of the dark matter. One popular model is thermal dark matter in which the dark matter is produced thermally in the early universe~\citep{2016PhLB..760..106B}. For a thermal relic WIMP, a velocity weighted cross-section of ${\langle\sigma_Av\rangle\cong \mbox{3} \times \mbox{10}^{-26}\rm\, cm^3 s^{-1}}$ in the early universe is needed in order to produce the dark matter density observed today. However, the kinematics of the dark matter today are very different than in the early universe. If the dark matter couples to gauge bosons, this can create a resonance which is amplified for low-velocity dark matter and significantly increases the dark matter cross-section with respect to thermal relic, a process referred to as Sommerfeld enhancement~\citep{ 2009PhRvD..79h3523L,sommerfeld}. Due to Sommerfeld enhancement, a cross-section $\langle\sigma_Av\rangle$ can be several orders of magnitude larger today compared to a cross-section of $\langle\sigma_Av\rangle$ in the early universe. In addition to Sommerfeld enhancement, other theoretical models also predict large dark matter cross-sections, particularly at masses $\gtrsim 10$\,TeV. Dark matter bound states, for example, can increase the dark matter cross-section to even higher cross-sections than Sommerfeld enhancement, approaching ${\langle\sigma_Av\rangle\sim 10^{-22}\rm\,  cm^3 s^{-1}}$~\citep{An:2016gad}.

In the TeV-PeV mass range, there has also been recent excitement about decaying dark matter. WIMP-like particles which decay may be responsible for the observation of an astrophysical neutrino excess by the IceCube detector~\citep{Aartsen:2013bka,Aartsen:2013jdh,Esmaili:2013gha,Bai:2013nga}. These particles could have large dark matter decay lifetimes and would produce gamma rays in similar quantity and energy to the observed neutrinos, from 100 TeV to several PeV~\citep{Kopp:2015bfa,Boucenna:2015tra}. The gamma-ray searches for this dark matter, as those in this paper, will provide additional information on  these possible high-flux, high-mass dark matter signals.

While there are many promising places in the Universe to look for signatures of dark matter, dwarf spheroidal galaxies (dSphs) are among the best candidates for a dark matter search. They are expected to be extremely dark matter rich, as the gravitational effects indicate much more mass present than the luminous material can account for. The dwarf spheroidal galaxies considered in this analysis are companion galaxies of the Milky Way, in what is known as our Local Group. They are very low luminosity galaxies, with low diffuse Galactic gamma-ray foregrounds and little to no astrophysical gamma-ray production~\citep{2015arXiv151000389B}. Due to these reasons, dSph can be used to probe the particle nature of dark matter (such as annihilation cross-section and decay lifetime).

While there are numerous dSphs near the Milky Way, a total of 15 are considered in this analysis: Bootes I, Canes Venatici I, Canes Venatici II, Coma Berenices, Draco, Hercules, Leo I, Leo II, Leo IV, Segue 1, Sextans, Ursa Major I, Ursa Major II, Ursa Minor and TriangulumII. These dSphs were chosen for their favorable declination angle for the HAWC observatory and well studied dark matter content.

In this paper, we calculate the expected gamma-ray flux due to annihilation and decay of dark matter for five channels. We search for dark matter gamma-ray signature from the 15 dSphs. Because no significant gamma-ray excess is observed, we report the corresponding upper limits for annihilation cross-section and lower limits for decay lifetime for 15 dSphs based on the calculated expected flux.

\section{HAWC Observatory}

The High Altitude Water Cherenkov (HAWC) observatory detects high-energy gamma-ray and is located at Sierra Negra, Mexico. The site is 4100 m above sea level, at latitude 18$^\circ$59.7' N and longitude 97$^\circ$18.6' W. HAWC is a survey instrument that is sensitive to gamma rays of 500 GeV to a few hundred TeV \citep{2017arXiv170101778A} energies. HAWC consists of 300 water Cherenkov detectors (WCDs) covering 22000 m$^2$ area. Each detector contains four photo-multiplier tubes \citep{2017arXiv170101778A}. It has been operating with a partial detector since August 2013 and has been operating with the full detector since March 2015. Here we present results from 507 days of its operations with the full detector.

Vast majority of the events detected by HAWC are cosmic rays. Gamma-hadron separation cuts are applied to remove the cosmic ray contamination from gamma ray events for different analysis bins (f$_{hit}$) which are defined by the fraction of the number of photomultiplier tubes (PMTs) in an event. Each of the analysis bins has a characteristic median energy for events but there is considerable overlap for consecutive analysis bins \citep{2017arXiv170101778A}.

\section{Dark Matter Gamma-ray Flux}
\subsection{Gamma-ray Flux from Dark Matter Annihilation}

A calculation of the expected gamma-ray flux from dark matter annihilation requires information about both the astrophysical properties of the potential dark matter source and the particle properties of the initial and final-state particles. The differential gamma-ray flux integrated over solid angle of the source is

\begin{equation} \label{Flux}
\frac{dF}{dE}_{annihilation} = \frac{\langle\sigma_{A}v\rangle}{8\pi M_{\chi}^{2}}\frac{dN_{\gamma}}{dE}J
\end{equation}
where $\langle\sigma_{A}v\rangle$ is the velocity-weighted dark matter annihilation cross-section, $dN_{\gamma}/dE$ is the gamma-ray spectrum per dark matter annihilation, and $M_{\chi}$ is the dark matter particle mass. $J$-factor ($J$) is defined as the dark mass density ($\rho$) squared and integrated along the line of sight distance $x$ and over the solid angle of the observation region
\begin{equation} \label{Jdeltaomega}
J= \int_{\rm source} d\Omega \int dx \rho^{2} (r(\theta,x)) 
\end{equation}
where the distance from the earth to a point within the source is given by
\begin{equation} \label{rgal}
r(\theta,x) = \sqrt{R^{2} - 2xR\cos(\theta) + x^{2}}\enspace,
\end{equation}
$R$ is the distance to the center of the source, and $\theta$ is the angle between the center of the source and the line of sight. 

\subsection{Gamma-ray Flux from Dark Matter Decay}

The gamma-ray flux from dark matter decay is similar to the dark matter annihilation gamma-ray flux as described above in Equation \ref{Flux}. The decay flux depends on the inverse of the dark matter lifetime $\tau$ instead of the annihilation cross-section.

\begin{equation} \label{Flux_decay}
\frac{dF}{dE}_{decay} = \frac{1}{4\pi \tau M_{\chi}}\frac{dN_{\gamma}}{dE}D\enspace.
\end{equation}

Because decays involve only one particle, not two, the gamma-ray flux from dark matter decay depends on a single power of the dark matter density $\rho$ instead of the square. This results in a D-factor for decay which differs from the annihilation J-factor and is given by:

\begin{equation}
D=\int_{\rm source} d\Omega \int dx \rho (r_{gal}(\theta,x))\enspace.
\end{equation}

Moreover, the total center of mass energy for each dark matter decay contains only half of the energy as an annihilation of two DM particles of similar masses.

\subsection{Dark Matter Density Distributions}\label{DMdenssec}

Density profiles describe how the density ($\rho$) of a spherical system varies with distance ($r$) from its center. In this paper, the Navarro-Frenk-White (NFW) model is used for the dark matter density profiles. The NFW profile~\citep{NFW, NFW2} is the simplest model consistent with N-body simulations. The NFW density profile is given by

\begin{equation} \label{NFW}
\rho_{\rm NFW} (r)= \frac{\rho_{s}}{(r/r_{s})^\gamma(1+(r/r_{s})^\alpha)^{(\beta-\alpha)/\gamma}}
\end{equation}
where $\rho_{s}$ is the scale density, $r_{s}$ is the scale radius of the galaxy, $\gamma$ is the slope for $r<<r_s$, $\beta$ is the slope for $r>>r_s$ and $\alpha$ is the transition parameter from inner slope to outer slope. The source parameter values for the 15 dwarf spheroidal galaxies presented are listed in Table \ref{table:sourceparameters}. The J-factor and D-factor for each source are calculated by the {\sc CLUMPY} software ~\citep{Bonnivard2016} using different realizations of the tabulated values and their respective uncertainties for an angular window of $\theta_{max}$. The median values for each source is given in Table \ref{table:sourceparameters}. Because of HAWC's angular resolution, we also calculated these parameters for integration angles of 0.2$^\circ$ and 1.0$^\circ$. For TriangulumII, these calculations were not performed due to the fact that the parameter set needed for calculations was not given in literature, so we use J- and D-factors by \citet{Hayashi:2016kcy}.

\subsection{Dark Matter Gamma-ray Spectra}

The {\sc Pythia} program models interactions between two incoming particles and their outgoing particles~\citep{Pythia}. This makes the program ideal for simulating interactions between two dark matter particles and monitoring the number of gamma rays we expect to see as a result of the dark matter annihilation. {\sc Pythia 8.2}~\citep{pythia8} was used in this analysis to calculate the expected photon spectrum for each WIMP annihilation channel. The photon radiation of charged particles was simulated, as well as the decay of particles such as the $\pi^{0}$~\citep{Abeysekara2014, ICRCpat}. For each annihilation channel and each dark matter mass, the average number of photons in energy bins per annihilation event was calculated. This differential flux, $dN_{\gamma}/dE$, was used to determine the dark matter gamma-ray flux of the targeted source. 

Due to the available phase space, dark matter will usually annihilate into the heaviest available channel, so we consider the heavy top quark ($t\bar t$) and  tau lepton ($\tau^{+}\tau^{-}$) channels. The bottom quark channel ($b\bar b$) is included since it has been studied by several experiments (Fermi-LAT, MAGIC, etc.) to allow for direct comparison of results. The W-channel ($W^{+}W^{-}$) was chosen since it is a standard bosonic channel that is widely considered in other experiments. Finally the muon channel ($\mu^{+}\mu^{-}$) is included in this analysis since dark matter models which are dominated by annihilation into light leptons may be able to explain measured excesses of local positrons~\citep{Cholis:2008wq}. For dark matter decay, the same channels are used.

\begin{figure}
\includegraphics[width=\columnwidth]{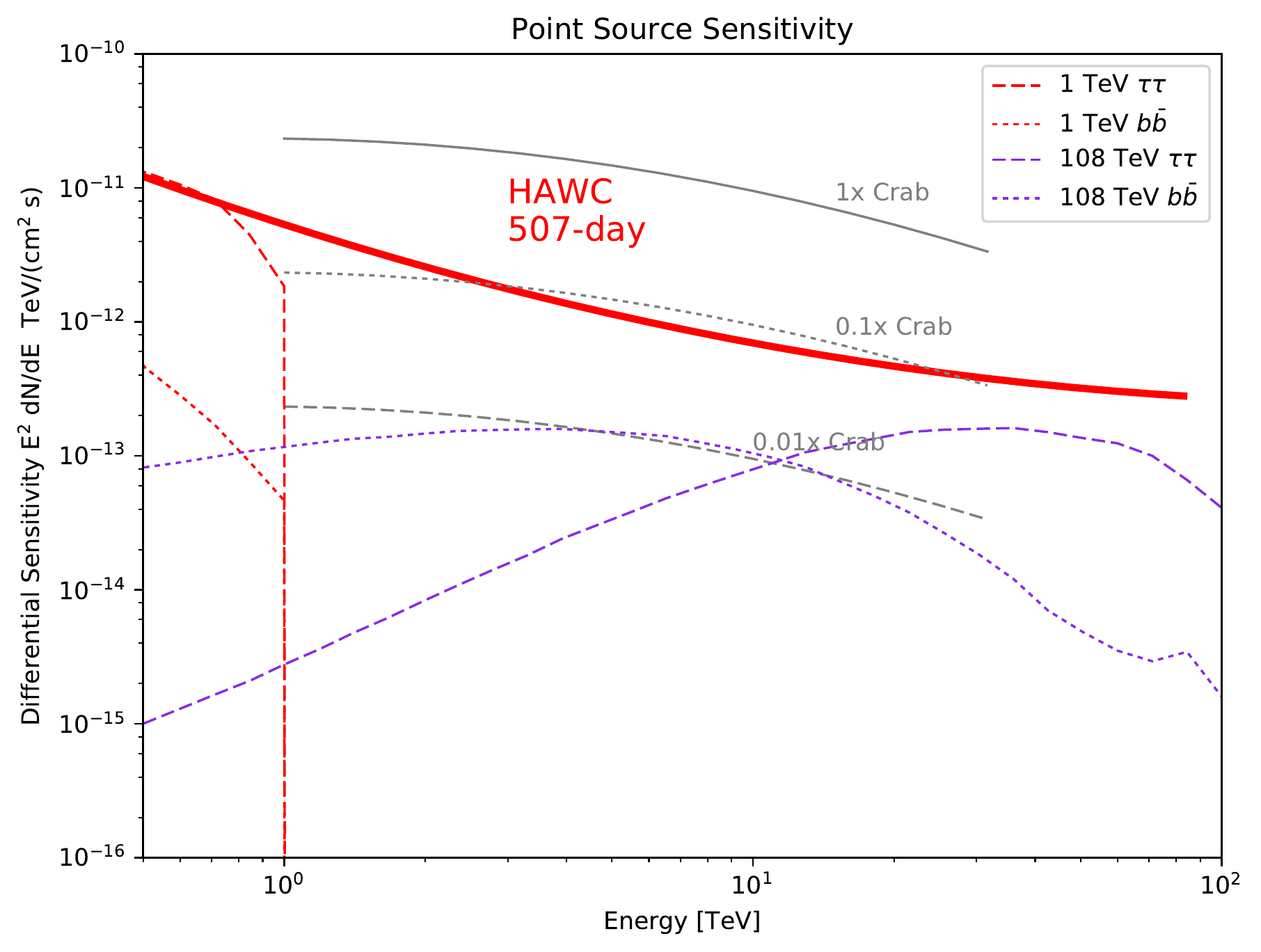} 
\caption{Expected gamma-ray flux from annihilation into two seperate channels ($\tau\tau$ and $b\bar{b}$) of 1 TeV and  108 TeV DM compared with HAWC point source sensitivity.}
\label{fig:pythiaExample}
\end{figure}

An example of the generated expected gamma-ray flux is shown in Figure \ref{fig:pythiaExample} for dark matter of mass 1 TeV and 108 TeV annihilating into $b\bar{b}$ and $\tau\tau$. In this work, we scanned dark matter masses from 1 TeV up to 100 TeV.

\section{Calculation of Limits on the Dark Matter Annihilation Cross-section and Decay Lifetime}\label{CalcSection}

To search for a gamma-ray excess in a particular region of the sky, we perform a likelihood ratio test. This allows us to calculate the significance of a source that has a low signal-to-noise ratio. For the likelihood of the signal region $\mathcal{L}$ and the likelihood of the background region $\mathcal{L}_0$, we calculate the test statistic 
\begin{equation} \label{likely}
TS = -2\ln\left(\frac{\mathcal{L}_0}{\mathcal{L}(S_{\rm max})}\right)
\end{equation}
where $S_{max}$ is the value of signal flux which maximizes the likelihood and minimizes $TS$. 

The process for setting 95\% CL limits consists of finding $S_{95}$, the amount of signal which would change the $TS$ value by 2.71~\citep{Fermi2014}, that is
\begin{equation}
TS(S_{\rm max}) - TS(S_{95}) = 2.71\enspace.
\end{equation}

For the purposes of our dark matter searches, the assumption that the null hypothesis is true is a good approximation, as we actually see little to no gamma-ray signal coming from the direction of the dwarf spheroidal galaxies. This can be seen in the sample significances given in Table~\ref{table:sourceparameters} and the plots in figures~\ref{fig:sigma_tautau} and~\ref{fig:sigma_rest1}. However, if the value of $S_{\rm max}$ is unphysical, i.e. $S_{\rm max}<0$, we set $\mathcal{L}(S_{\rm max}) = \mathcal{L}_0$ so that $TS(S_{\rm max})$ is replaced by $-2\ln\left({\mathcal{L}_0}/{\mathcal{L}(S_{\rm max}}\right) = 0$ and instead and we find $S_{95}$ by
\begin{equation}
 - TS(S_{95}) = 2.71\enspace.
\end{equation}

After having determined the allowed amount of signal flux at 95\% CL, we solve equation~\ref{Flux} or equation~\ref{Flux_decay} to find the corresponding values $\langle\sigma_{A}v\rangle_{95}$ and $\tau_{95}$ which produce that flux.

The joint likelihood analysis is a stacked study of many dwarf spheroidal galaxies. A combined analysis increases the overall statistical power and produces a better constraint on the dark matter annihilation cross-section and decay lifetime. The same likelihood analysis procedure is followed as described in the above section. However, the likelihood values are instead summed over all sources rather than over a single source. For a more detailed discussion of the limit-setting procedure, see Appendix~\ref{CalcAppendix}.

\begin{longrotatetable}
  \begin{deluxetable*}{ccccccccccccc}
    \tablewidth{900pt}
    \tablecaption{Astrophysical parameters, J-factors and D-factors, for the fifteen dwarf spheroidal galaxies within the HAWC field-of-view. The source, right ascension ($RA$), declination ($Dec$), scale density ($\rho_{s}$), scale radius ($r_{s}$), distance to the source ($R$), the dark matter $J$-factor and $D$-factor are listed above. NFW profiles parameters from~\cite{Geringer-Sameth2015} are used for all sources listed, except for Triangulum II. For Triangulum II, we use the $J$ and $D$ factors from~\cite{Hayashi:2016kcy}. J-factors and D-factors are calculated for integration angle of $\theta_{max}$ for respective sources. The significance ($\sigma$) is also shown for each source for $M_{\chi} = 12~\rm\, TeV$ and the $\chi\chi\rightarrow \tau^+\tau^-$ annihilation channel.}
    \label{table:sourceparameters}
    \centering
    \tabletypesize{\small}
    \tablehead{
      \colhead{Source} &
      \colhead{RA} &
      \colhead{Dec} &
      \colhead{Distance} &
      \colhead{log$_{10}$ ($\rho_s$/$M_{\odot}/pc^3$)} &
      \colhead{log$_{10}$ ($r_s$/pc)} &
      \colhead{$\theta_{max}$} &
      \colhead{$\alpha$} &
      \colhead{$\beta$} &
      \colhead{$\gamma$} &
      \colhead{log$_{10}$ (J/$GeV^2 cm^{-5} sr$)} &
      \colhead{log$_{10}$ (D/$GeV cm^{-5} sr$)} &
      \colhead{$\sqrt{TS}$} \\
      \colhead{} &
      \colhead{[deg]} &
      \colhead{[deg]} &
      \colhead{[kpc]} &
      \colhead{} &
      \colhead{} &
      \colhead{} &
      \colhead{} &
      \colhead{} & 
      \colhead{} & 
      \colhead{} & 
      \colhead{} & 
      \colhead{[$\sigma$]}
    }
    \startdata
      Bootes1         & 210.05 & 14.49 & 66  & -1.74 & 3.80 & 0.47 & 1.96 & 5.91 & 0.53 & 18.47 & 18.45 & -0.43 \\ 
      CanesVenaticiI  & 202.04 & 33.57 & 218 & -1.87 & 3.36 & 0.53 & 1.86 & 6.00 & 0.67 & 17.62 & 17.55 & -0.11 \\
      CanesVenaticiII & 194.29 & 34.32 & 160 & -1.46 & 3.91 & 0.13 & 1.92 & 6.09 & 0.43 & 17.95 & 17.68 & -0.50 \\ 
      ComaB           & 186.74 & 23.90 & 44  & -0.99 & 3.71 & 0.31 & 1.99 & 6.44 & 0.43 & 19.32 & 18.71 & -1.35 \\ 
      Draco           & 260.05 & 57.07 & 76  & -1.74 & 3.57 & 1.30 & 2.01 & 6.34 & 0.71 & 19.37 & 19.15 & 0.37  \\
      Hercules        & 247.72 & 12.75 & 132 & -1.62 & 2.93 & 0.28 & 1.81 & 6.39 & 0.55 & 16.93 & 16.87 & -1.05 \\
      LeoI            & 152.11 & 12.29 & 254 & -2.18 & 3.80 & 0.45 & 1.93 & 6.15 & 0.84 & 17.57 & 18.04 & -0.40 \\
      LeoII           & 168.34 & 22.13 & 233 & -0.92 & 2.89 & 0.23 & 1.76 & 5.95 & 0.82 & 18.11 & 17.33 & 0.17  \\
      LeoIV           & 173.21 & -0.53 & 154 & -1.80 & 2.94 & 0.16 & 1.93 & 6.32 & 0.55 & 16.37 & 16.50 & -0.16 \\
      Segue1          & 151.75 & 16.06 & 23  & -1.06 & 3.26 & 0.35 & 1.91 & 6.39 & 0.76 & 19.66 & 18.64 & -0.13 \\
      Sextans         & 153.28 & -1.59 & 86  & -2.39 & 3.79 & 1.70 & 1.98 & 6.06 & 0.63 & 17.96 & 18.59 & 0.43  \\
      TriangulumII    & 33.32  & 36.18 & 30  & -     & -    & -    & -    & -    & -    & 20.44 & 18.42 & -0.46 \\
      UrsaMajorI      & 158.72 & 51.94 & 97  & -1.84 & 3.50 & 0.53 & 1.87 & 6.25 & 0.71 & 18.66 & 18.11 & -0.03 \\
      UrsaMajorII     & 132.77 & 63.11 & 32  & -1.13 & 3.63 & 0.43 & 1.86 & 6.37 & 0.58 & 19.67 & 19.05 & 0.32  \\
      UrsaMinor       & 227.24 & 67.24 & 76  & -0.50 & 2.60 & 1.37 & 1.64 & 5.29 & 0.78 & 19.24 & 17.92 & -0.30 \\
      \enddata
  \end{deluxetable*}
\end{longrotatetable}

\section{Limits on the Dark Matter Annihilation Cross Section and Decay Lifetime with HAWC data}

\begin{figure}
  \includegraphics[width=.48\textwidth]{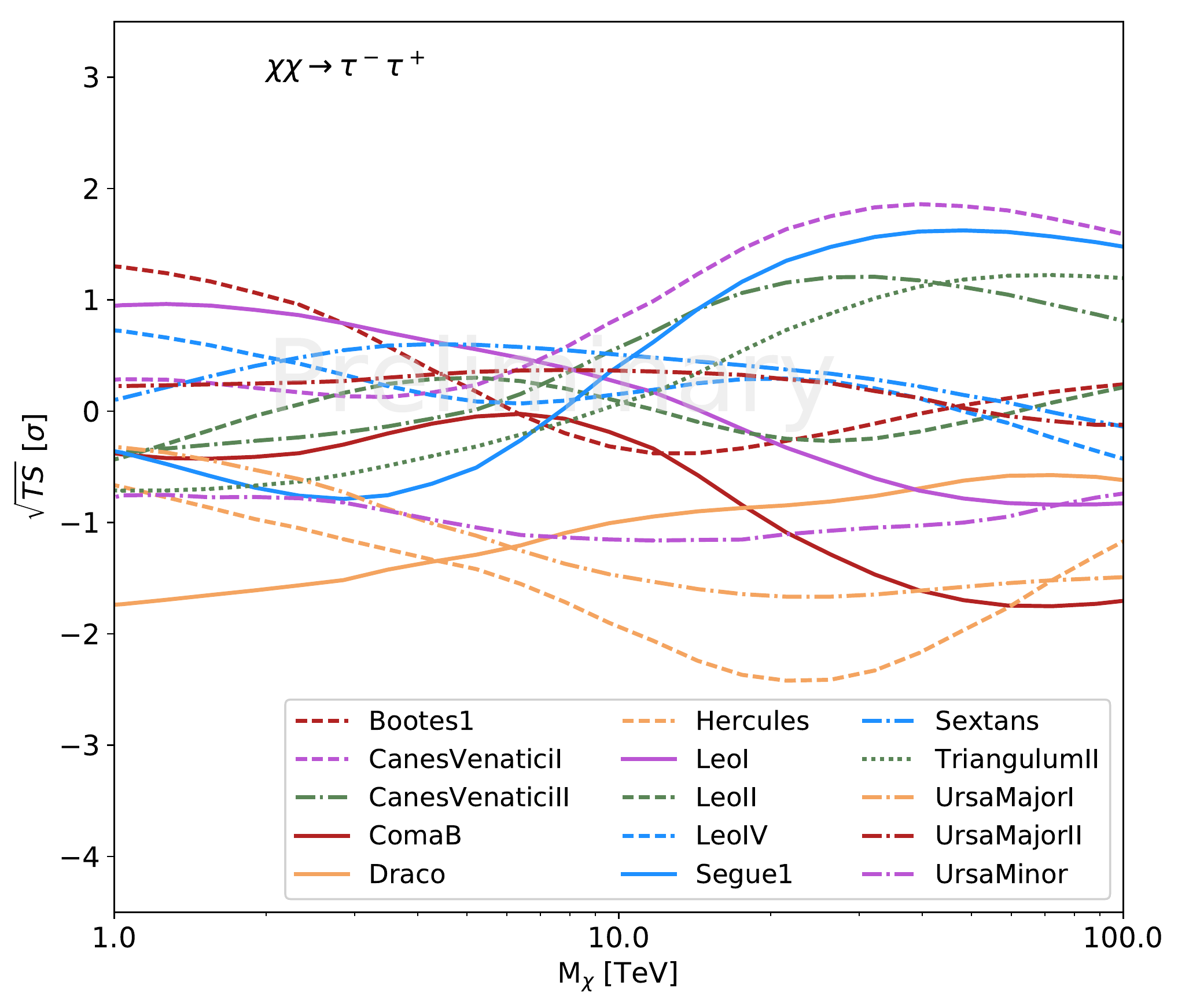}
  \caption{Statistical significance of dark matter annihilating into $\tau\bar{\tau}$ channel for the selected sources}
  \label{fig:sigma_tautau}
\end{figure}

\begin{figure*}[h]
  \includegraphics[width=.48\textwidth]{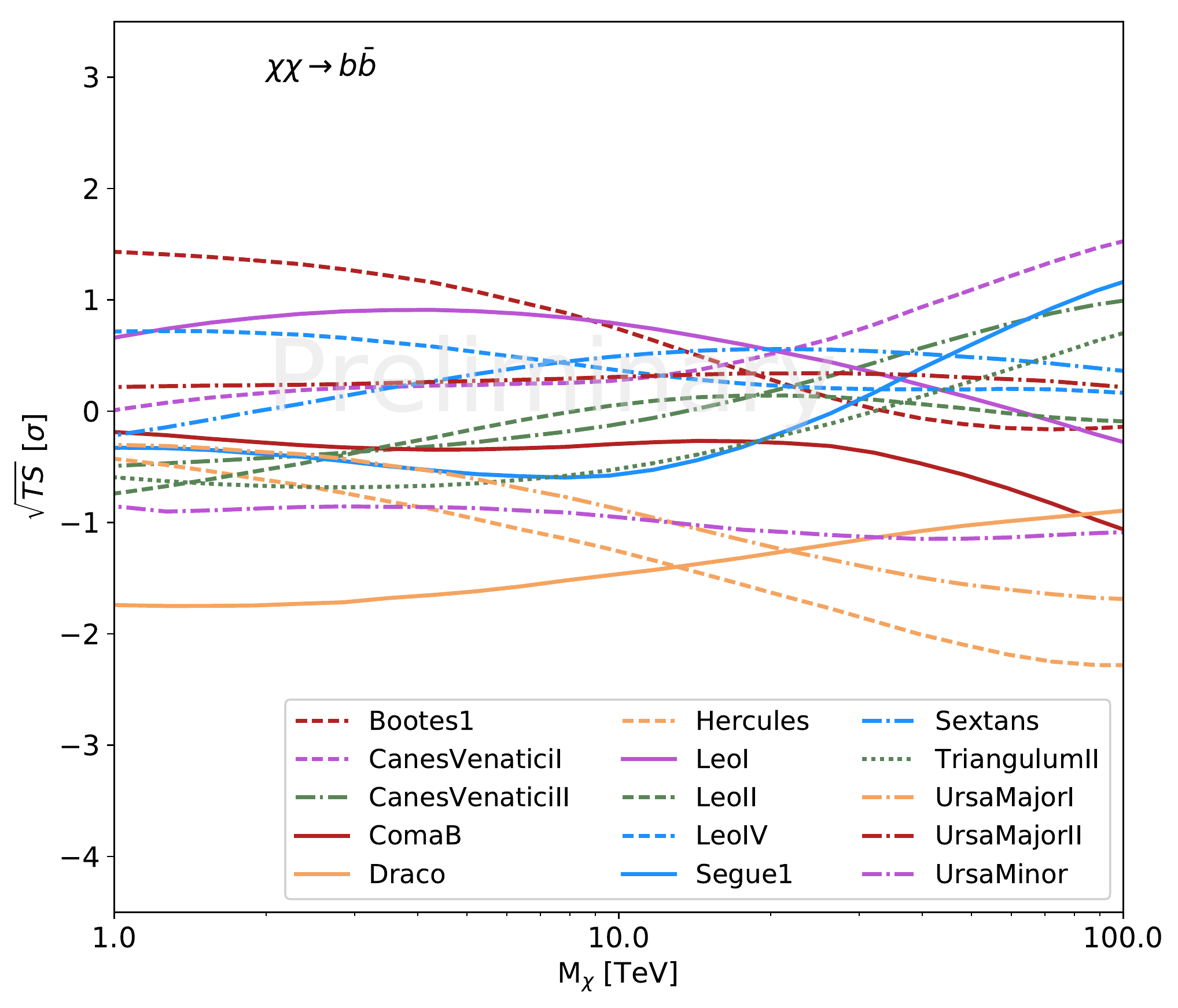}
  \includegraphics[width=.48\textwidth]{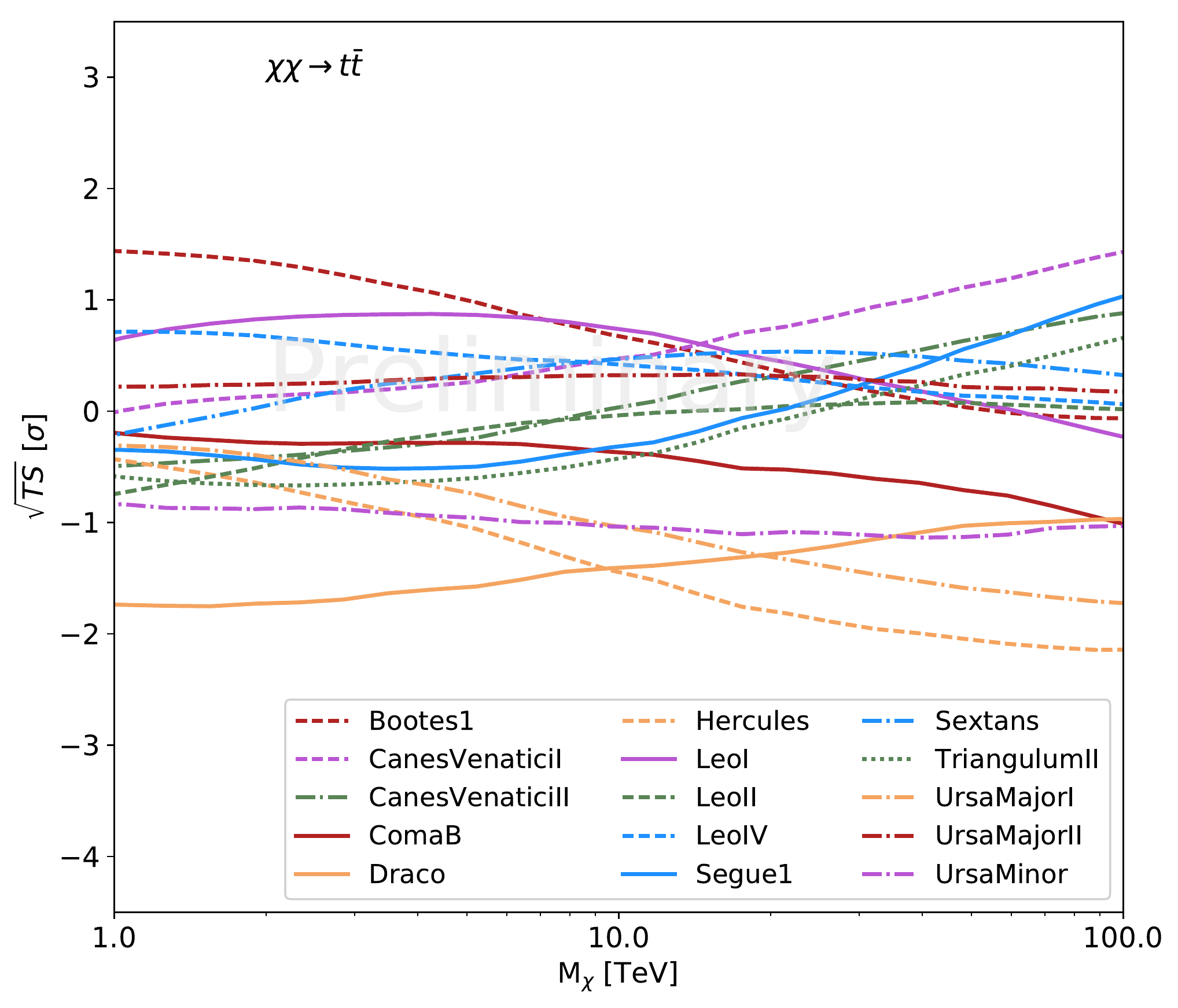}\\
  \includegraphics[width=.48\textwidth]{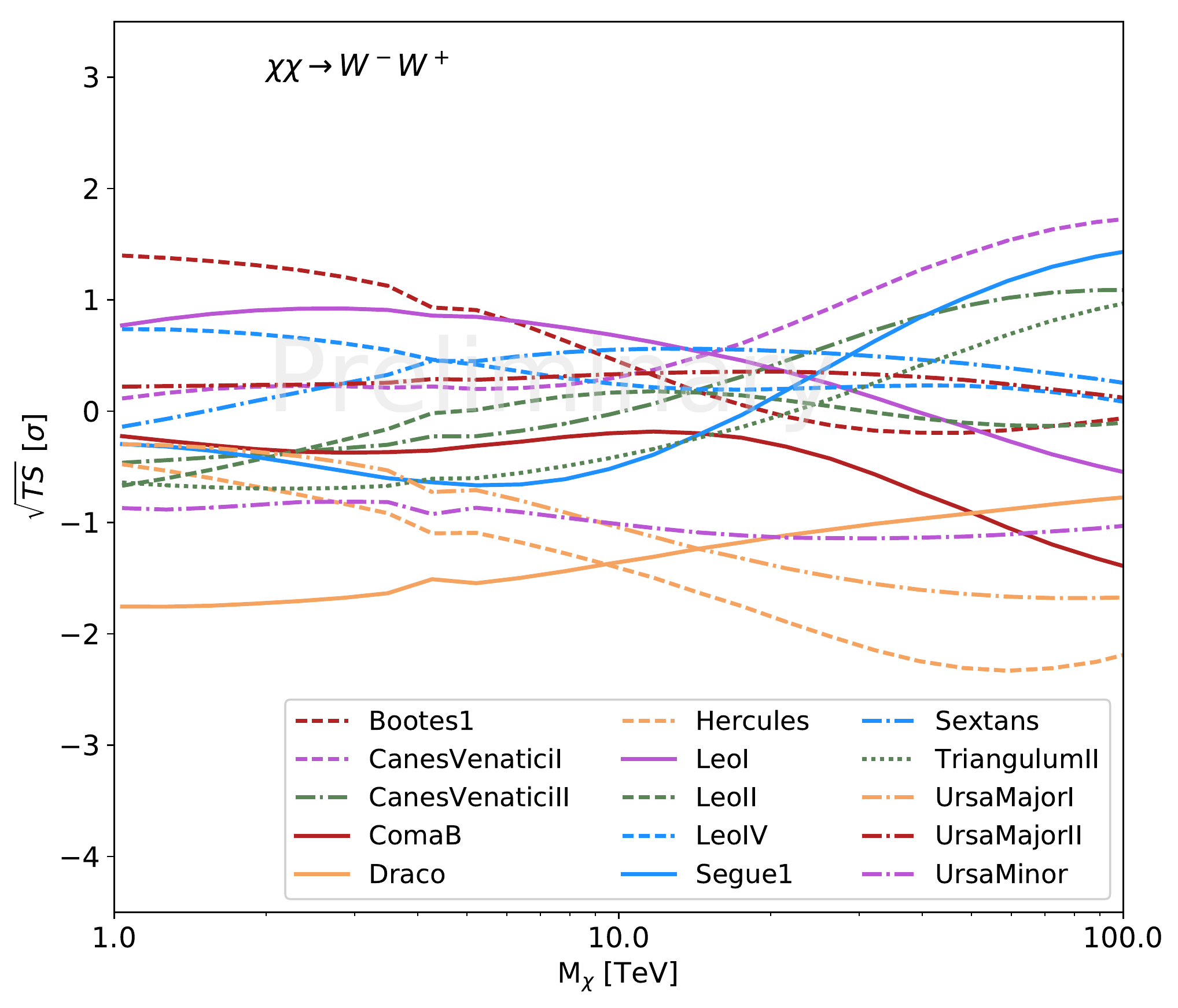}
  \includegraphics[width=.48\textwidth]{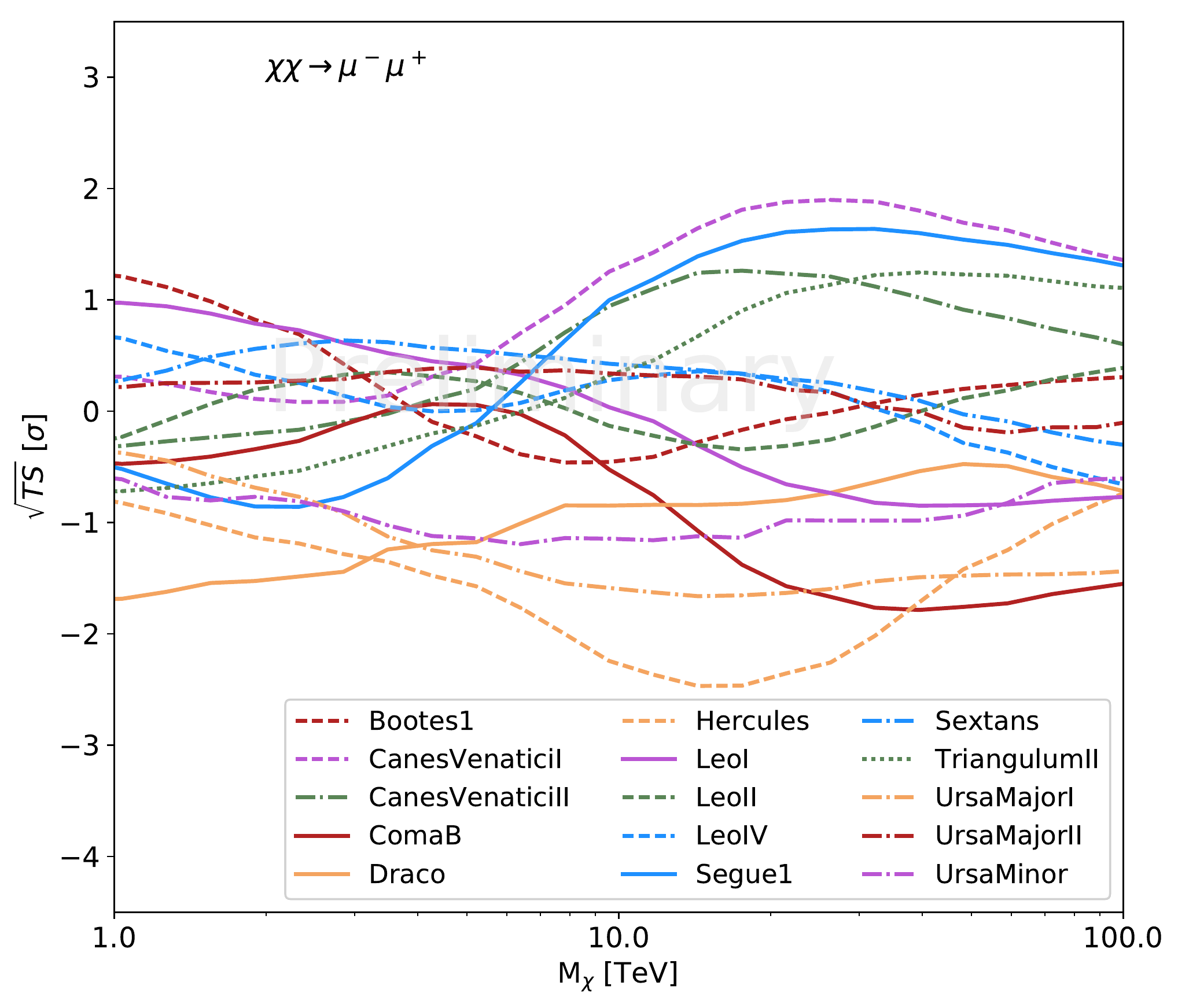}
  \caption{Statistical significance of dark matter annihilating into $b\bar{b}$, $t\bar{t}$, $W^-W^+$ and $\mu^-\mu^+$ channels for the selected sources}
  \label{fig:sigma_rest1}
\end{figure*}

Presented in this analysis are individual and combined limits from 15 dwarf spheroidal galaxies within the HAWC field of view for the HAWC 507 days data. Considering the angular resolution of HAWC observatory ($\sim$0.5 degrees) \citep{2017arXiv170101778A}, the limits were calculated assuming that the dSphs are point sources. Through detailed simulation of the HAWC gamma-ray sensitivity and backgrounds, the significance of the gamma-ray flux for a range of dark matter masses, 1~TeV - 100~TeV, and five dark matter annihilation channels has been found. In Figures \ref{fig:sigma_tautau} and \ref{fig:sigma_rest1}, we show the significance of dark matter annihilation into the selected channels. Since no significant gamma-ray excess was observed, $95\%$ confidence level limits were placed on the annihilation cross-section and decay lifetime using the method described in Section \ref{CalcSection}, the source significance is used to determine the exclusion curves on the dark matter annihilation cross-section $\langle \sigma_{A}v\rangle$ and decay lifetime $\tau$, for the individual dSphs. A joint likelihood analysis was also completed by combining the statistics for all 15 dSphs in order to increase the sensitivity of the analysis.

Triangulum II has a particularly large $J$-factor and transits near the zenith for HAWC. However, this dwarf was only discovered recently~\citep{2015ApJ...802L..18L} and still has large uncertainties in its mass profile. Because of this, we show the joint dwarf limit both including and excluding Triangulum II in Figures~\ref{15dwarffigs_annihilation} and~\ref{fig:15dwarffigs_decay}.

It should be noted that the ability to include data collected on a newly discovered dwarf galaxy since the beginning of the experiment is a strength of the HAWC design. With the HAWC wide field of view and large uptime, data for all declinations within the HAWC field of view exist, so further dwarf galaxies can be added to this analysis once their position and dark matter content are known.

\subsection{Systematics}

Systematic uncertainties arise from a number of sources within the detector, for example the uncertainties associated from taking data at different stages of the detector. Since HAWC was operational during its construction, there are data uncertainties due to the changing number of online WCDs and PMTs, the effects of which were studied using different simulations assuming a different number of PMTs. This effect is minimal considering the dataset used in the analysis, but it is included in the systematics uncertainties. In addition, the difference among signal passing rates was compared for simulations varying detector parameters within their systematic errors. An uncertainty comes from the measured number of photo-electrons (PEs) based on how well we simulate the detector, since muon studies have shown there is a discrepancy between the simulated PMT charge and the charge from actual data. There is also an uncertainty associated with the angular resolution of HAWC. Also, relative PMT photon detection efficiency and charge resolution vary from PMT to PMT. 

With these effects taken into consideration, \citet{2017arXiv170101778A} found that this gives an overall systematic uncertainty on the HAWC data set on the order of $\pm50\%$ on the observed flux. The uncertainties on the expected dark matter annihilation and decay limits were calculated to account for these systematics uncertainties.

There are also systematic uncertainties on the expected dark matter flux due to the integration angle of J- and D-factor. HAWC's angular resolution changes between 1$^\circ$ and 0.2$^\circ$ for near zenith angles. For better angular resolutions, the integration angle gets smaller that results in smaller J- and D-factors. Similarly, for worse angular resolutions, the integration angle gets greater that returns in larger values. However, there is a physical constraint on the dark matter distribution which limits the dark matter content of a source at angles larger than $\theta_{max}$. We impose this physically motivated constraint on the J- and D- factor uncertainties, resulting in a one-side uncertainty. For combined limit uncertainties, we used the uncertainties corresponding to Segue1 (42\% for annihilation cross-section limits and 38\% for decay lifetime limits) since it is one of the strongest sources that is driving the limits. Under current context, it would have been better to calculate and use this uncertainties for TriangulumII but the required information is not available for now.

\subsection{Flux Upper Limits}\label{flimits}

The annihilation cross-section and decay lifetime results are dependent on the dark matter annihilation or decay gamma ray flux used for the sources. A quasi-model independent upper flux limit for each source can be calculated to provide data for testing other models, following the same method outlined by \citet{2017arXiv170206131A}. We calculated flux limits separately for five energy bins with width 0.5log(E/TeV) centered at 1 TeV, 3.16 TeV, 10TeV, 31.6 TeV and 100 TeV, assuming a flux that is non-zero only within a given interval. 95 \% confidence limits were calculated for each energy bin assuming the flux is a power law with a spectral index of $\Gamma$=-2. The limits were also tested assuming spectral indices between 0 and -3 to study the spectral index dependence. The results were consistent within the systematic HAWC flux uncertainties~\citep{2017arXiv170101778A}. We report the normalization factors of the power law with spectral index of $\Gamma$=-2 for the upper flux limits in Table~\ref{tbl:upperflux}. For a discussion on using these limits, see Appendix~\ref{LimitAppendix}.

\subsection{Dark Matter Annihilation Cross-section Limits}

\begin{figure*}[h]
  \includegraphics[width=0.48\textwidth]{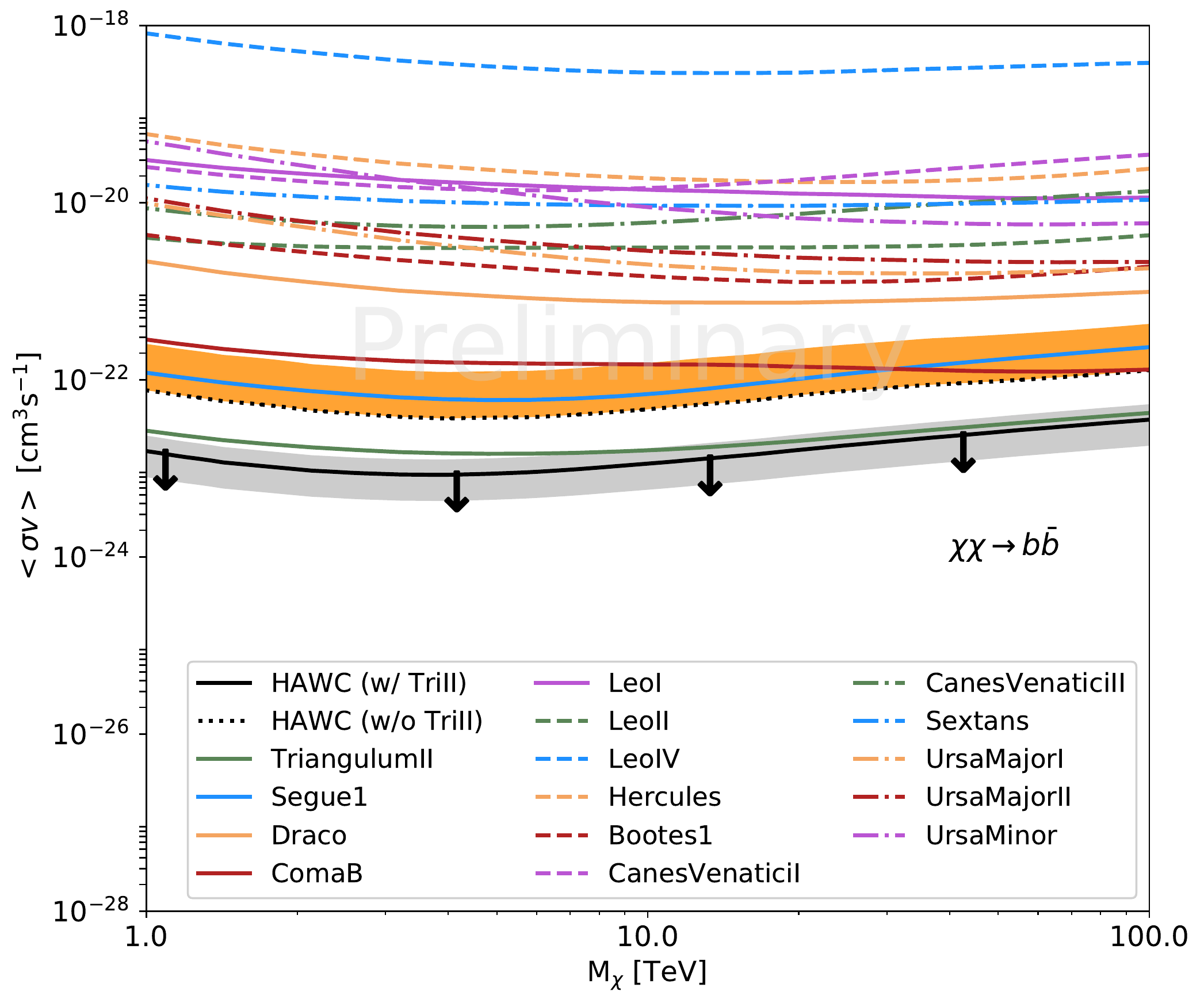}
  \includegraphics[width=0.48\textwidth]{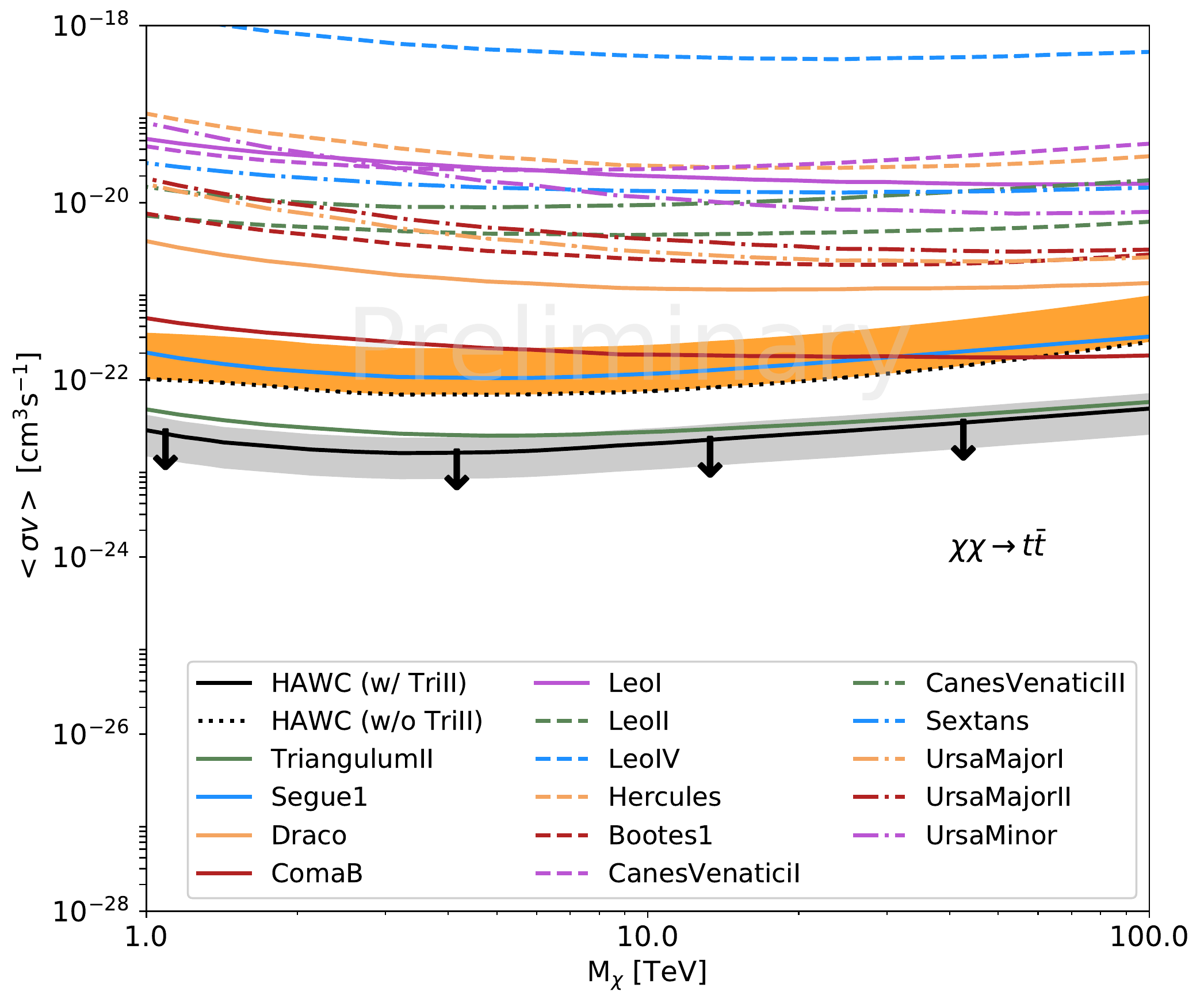}\\
  \includegraphics[width=0.48\textwidth]{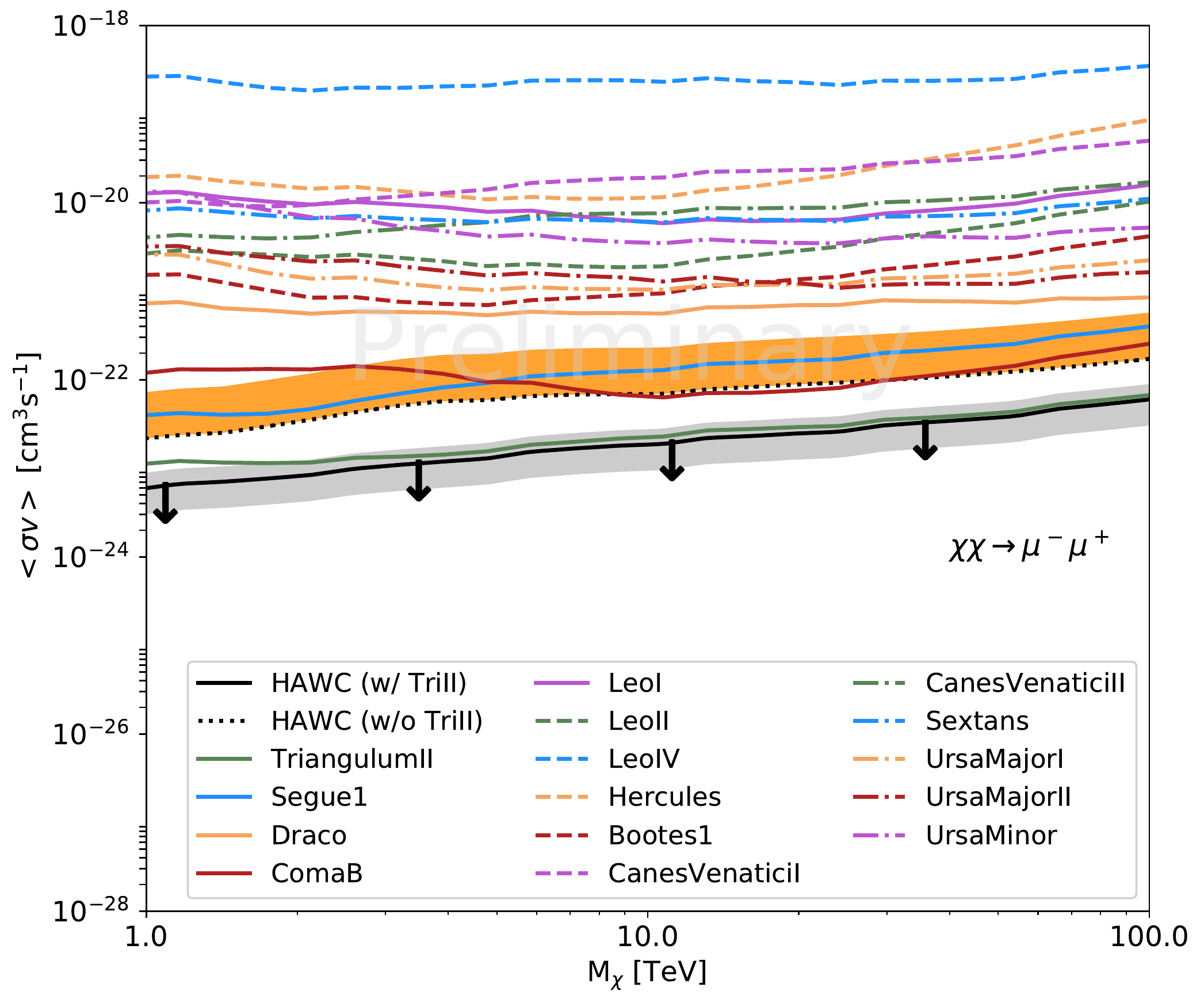}
  \includegraphics[width=0.48\textwidth]{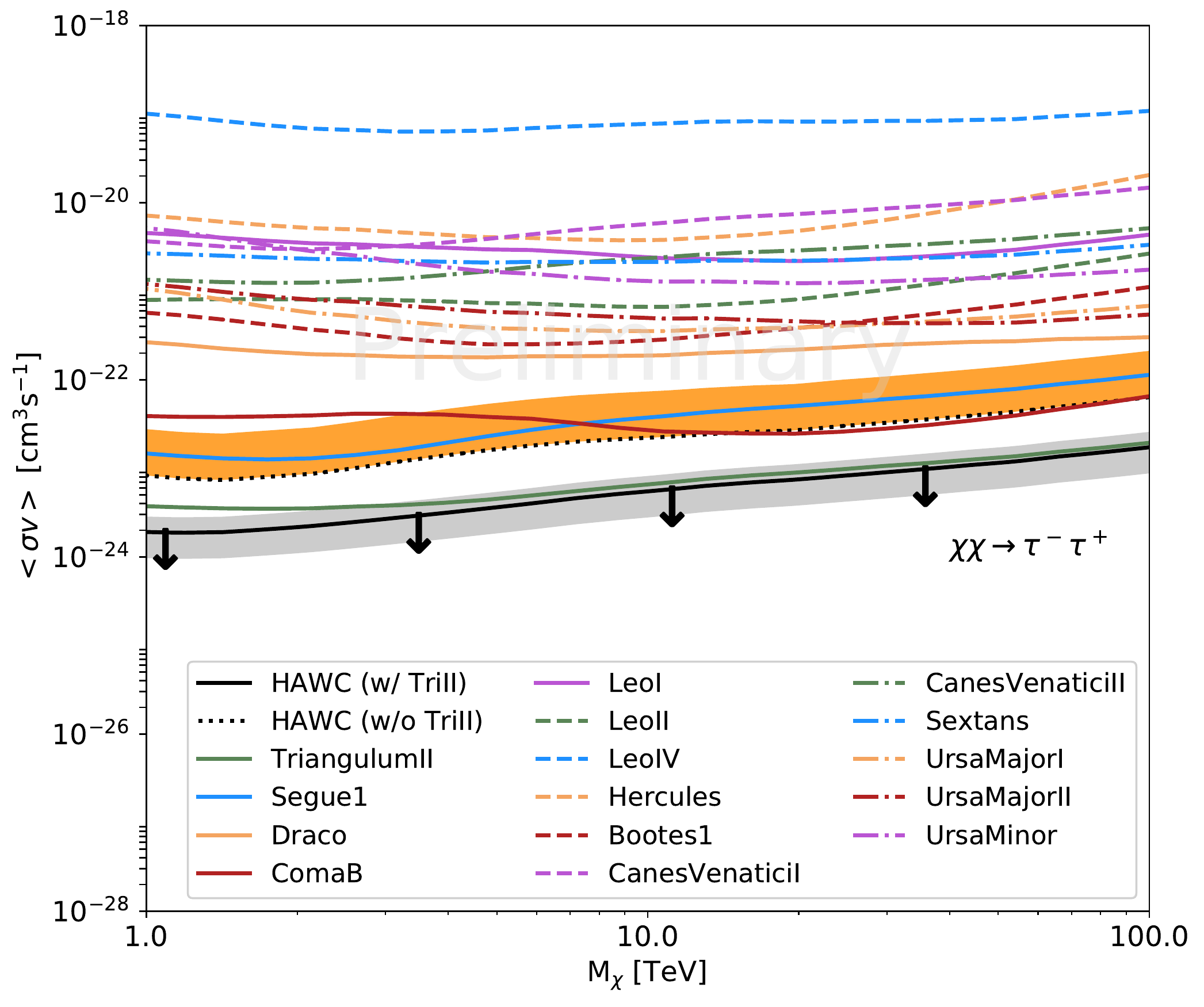}\\
  \includegraphics[width=0.48\textwidth]{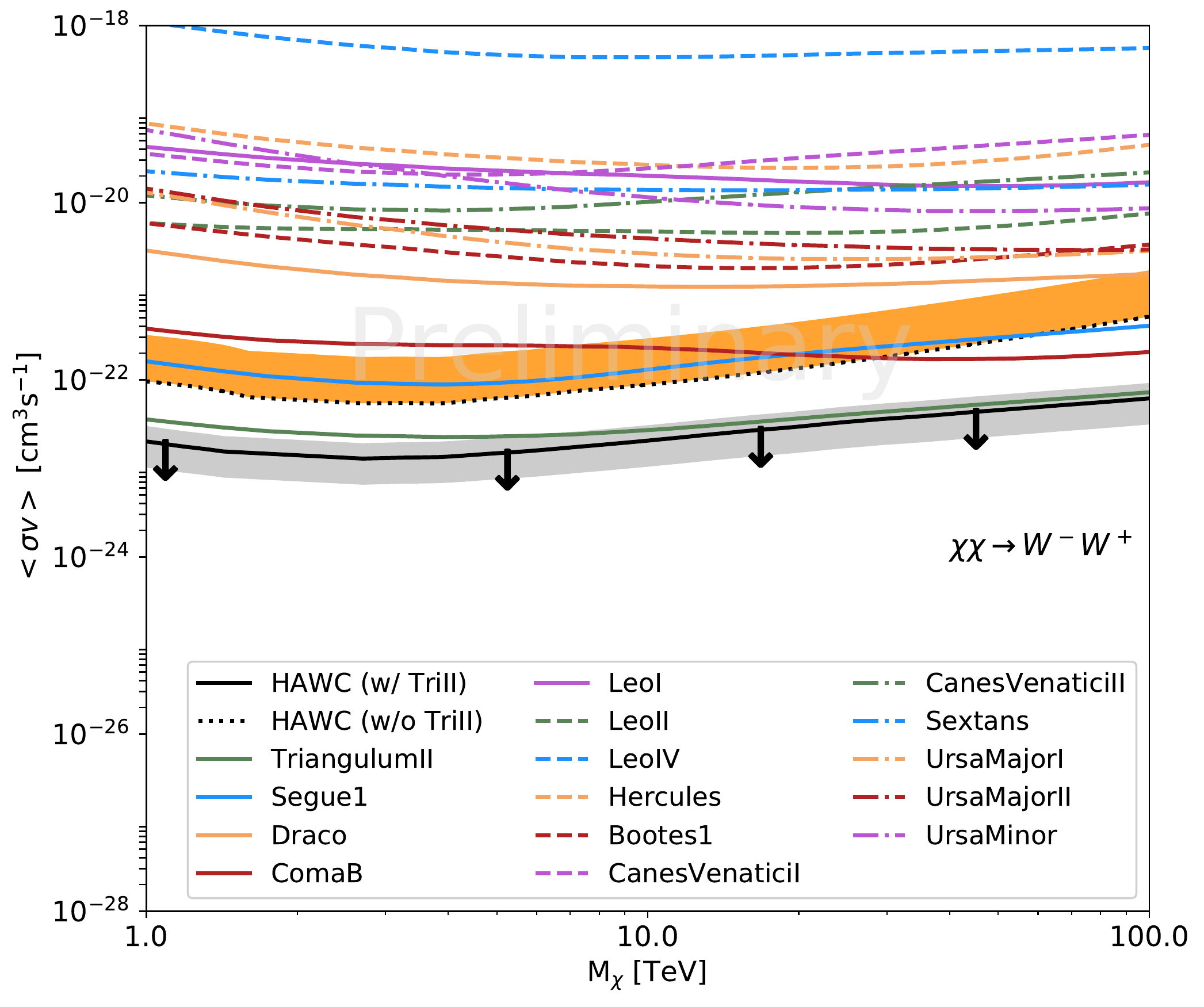}
  \caption{95$\%$ confidence level upper limits on the dark matter annihilation cross-section for 15 dwarf spheroidal galaxies within the HAWC field of view for the $b\bar{b}$, $t\bar{t}$, $\tau^{+}\tau^{-}$, $\mu^{+}\mu^{-}$ and $W^{+}W^{-}$ annihilation channels. The solid black line shows the combined limit using all dSphs resulting from a joint likelihood analysis. The dashed black line shows the combined limit using 14 dSphs, excluding Triangulum II. The gray band shows the systematic uncertainty on the combined limits due to HAWC systematics and dark orange band shows the systematic uncertainty due to J-factor uncertainty.
\label{15dwarffigs_annihilation}}
\end{figure*}

\begin{figure*}[h]
  \includegraphics[width=.46\textwidth]{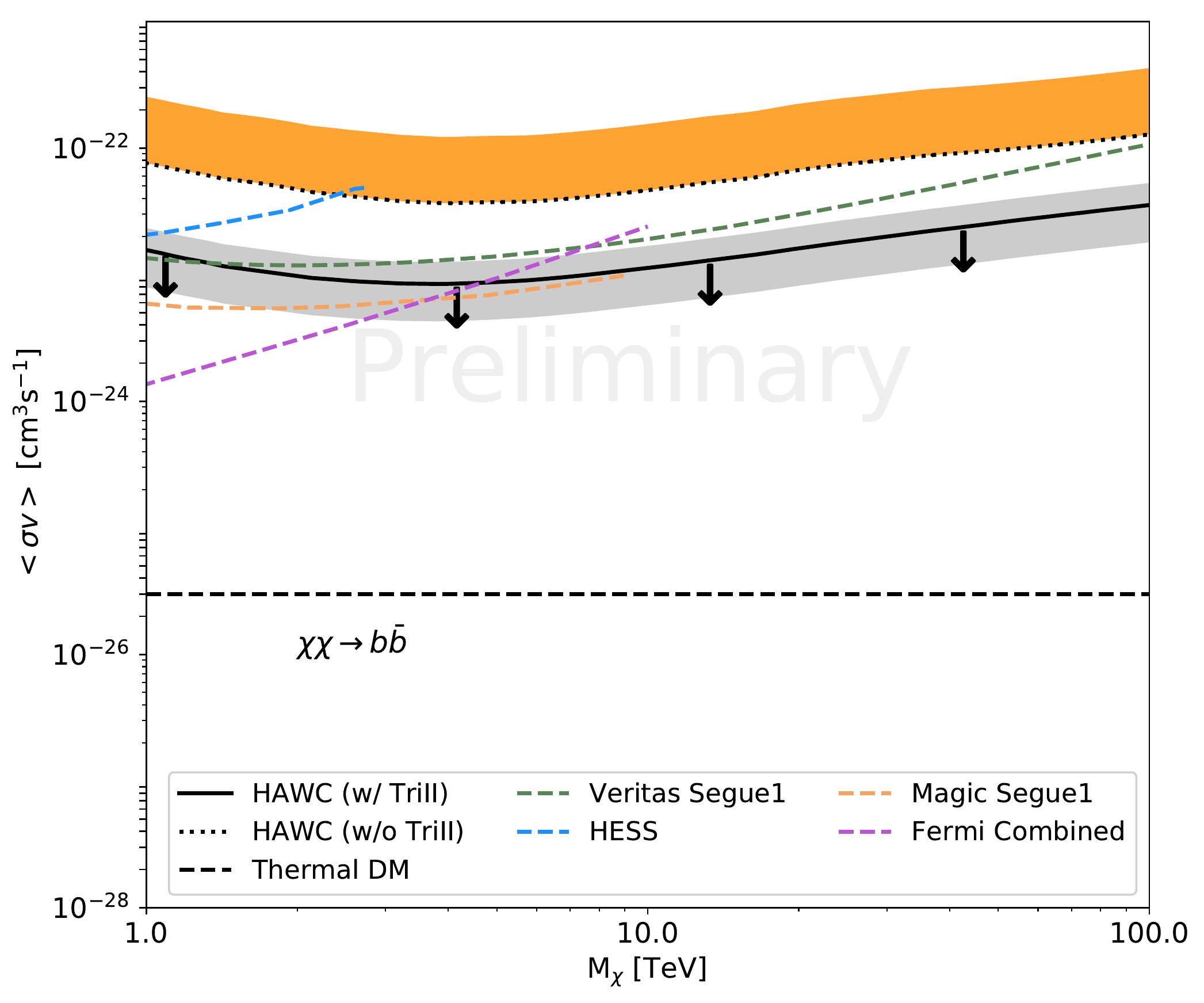}
  \includegraphics[width=.46\textwidth]{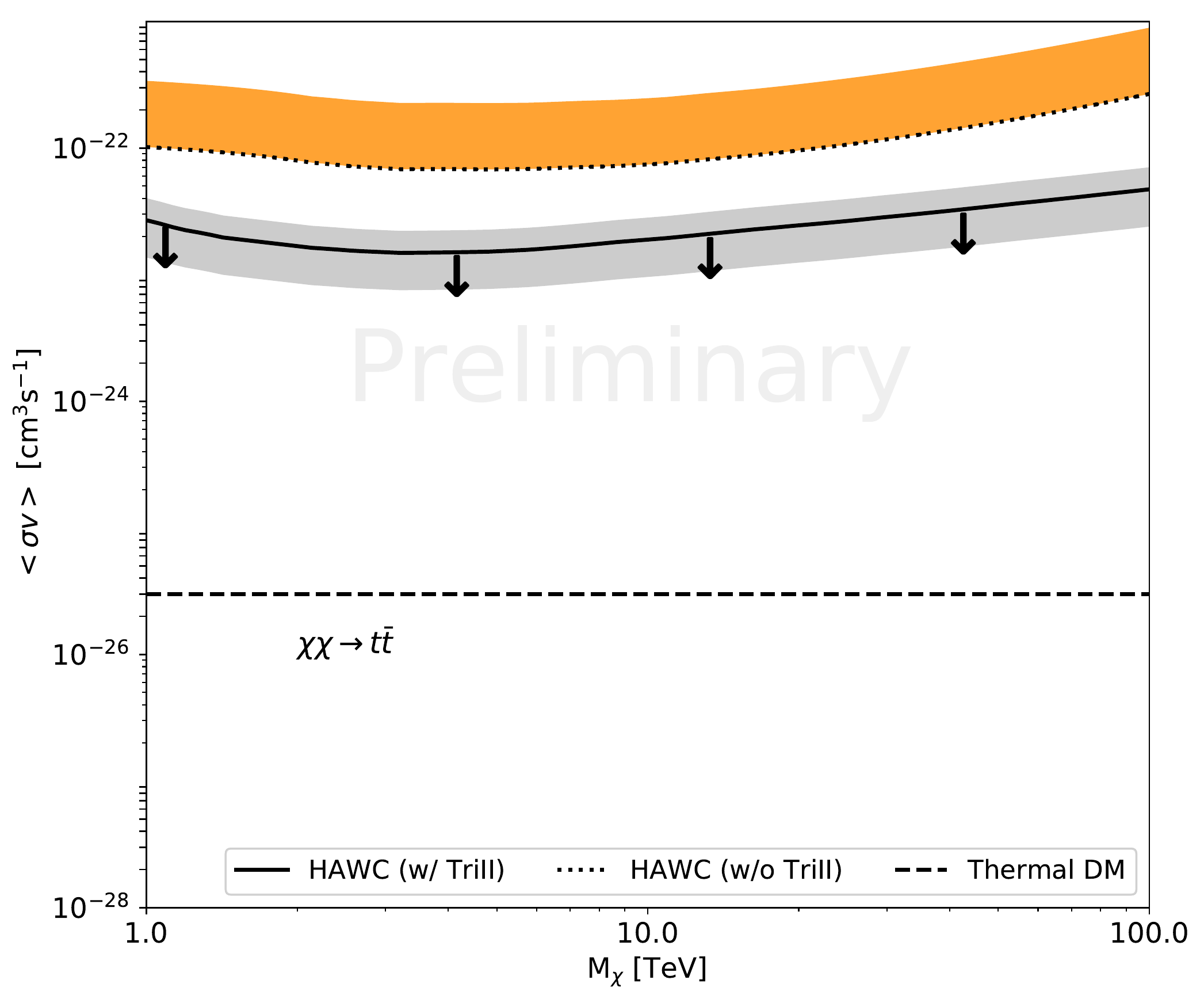}\\
  \includegraphics[width=.46\textwidth]{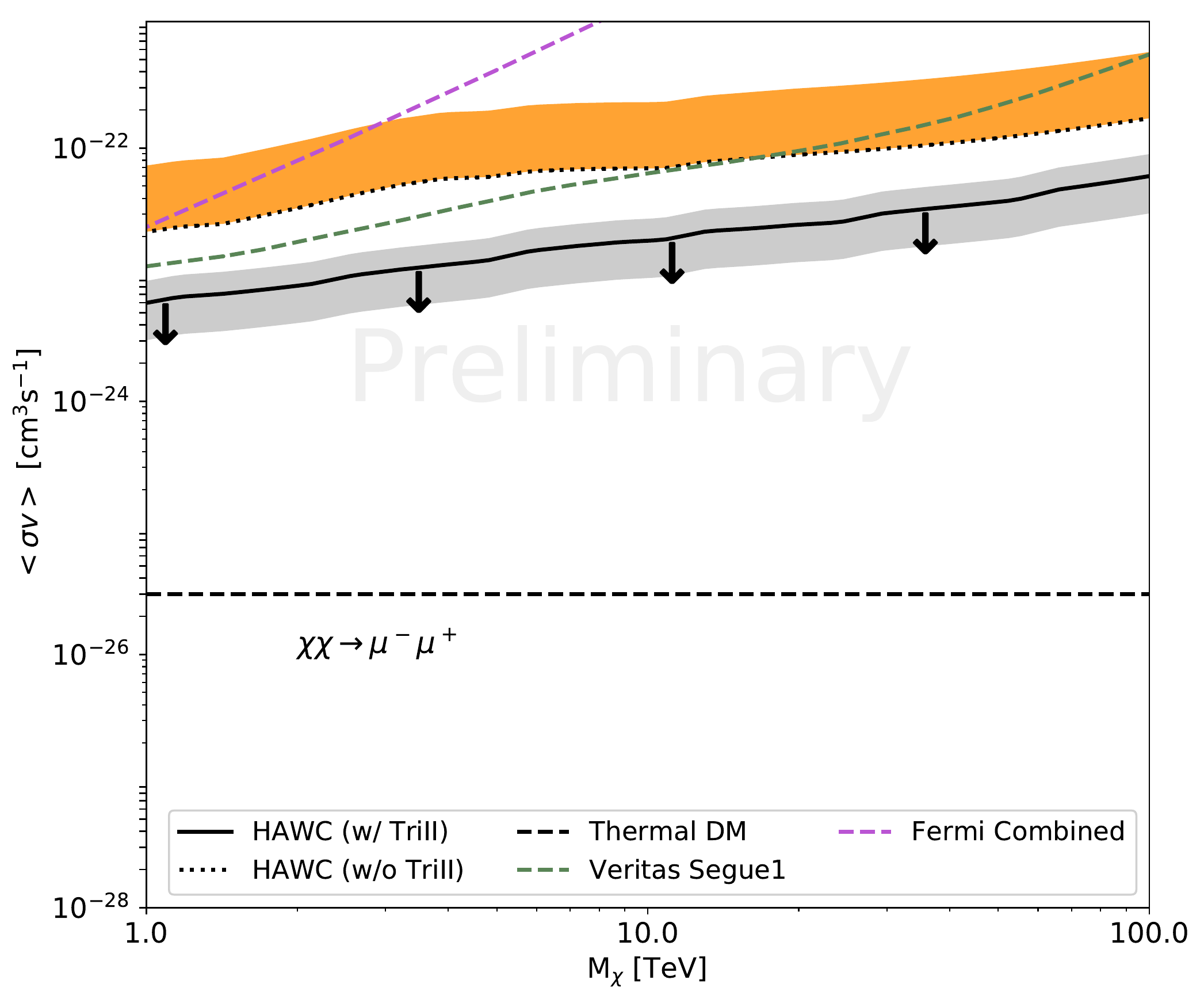}
  \includegraphics[width=.46\textwidth]{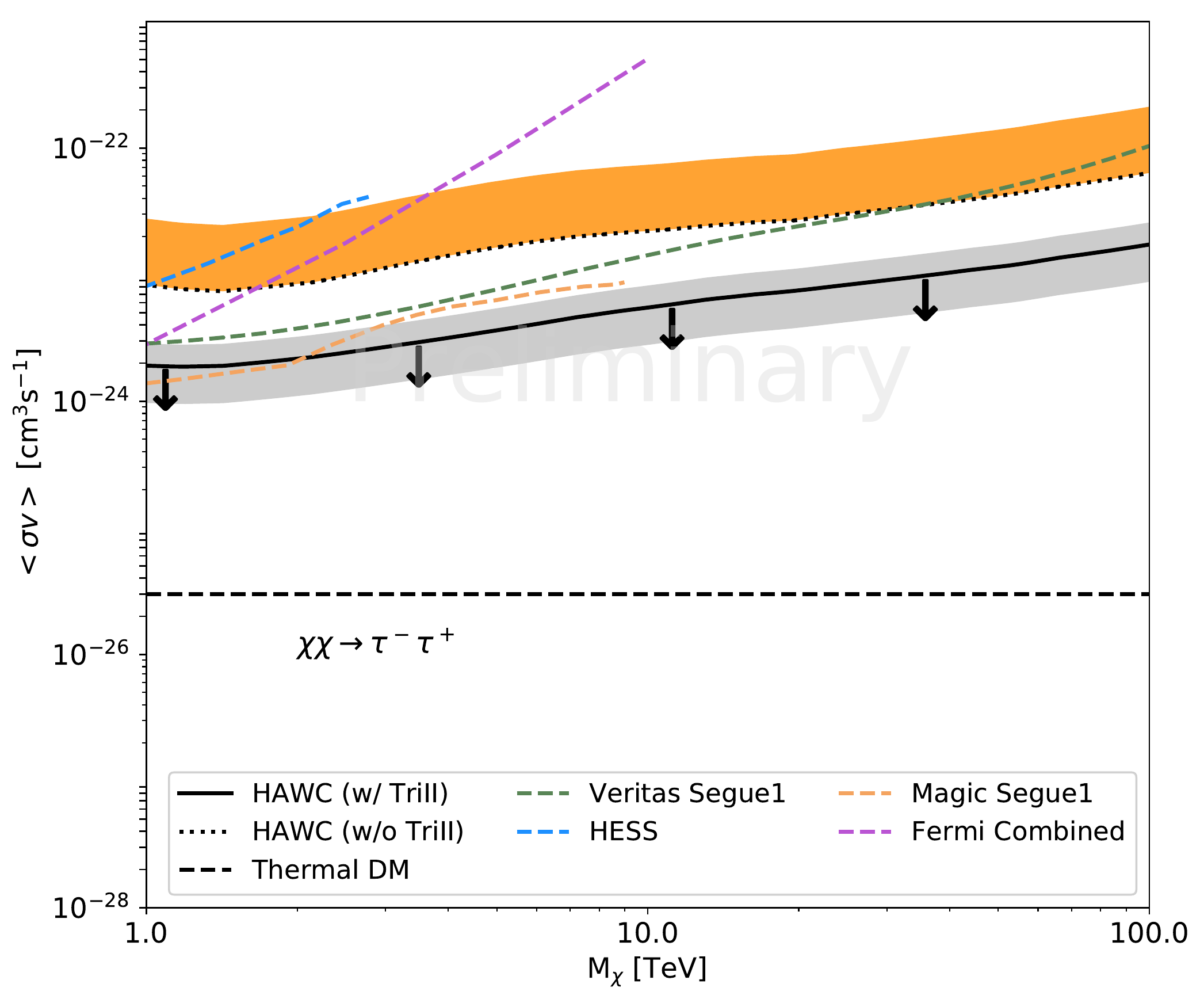}\\
  \includegraphics[width=.46\textwidth]{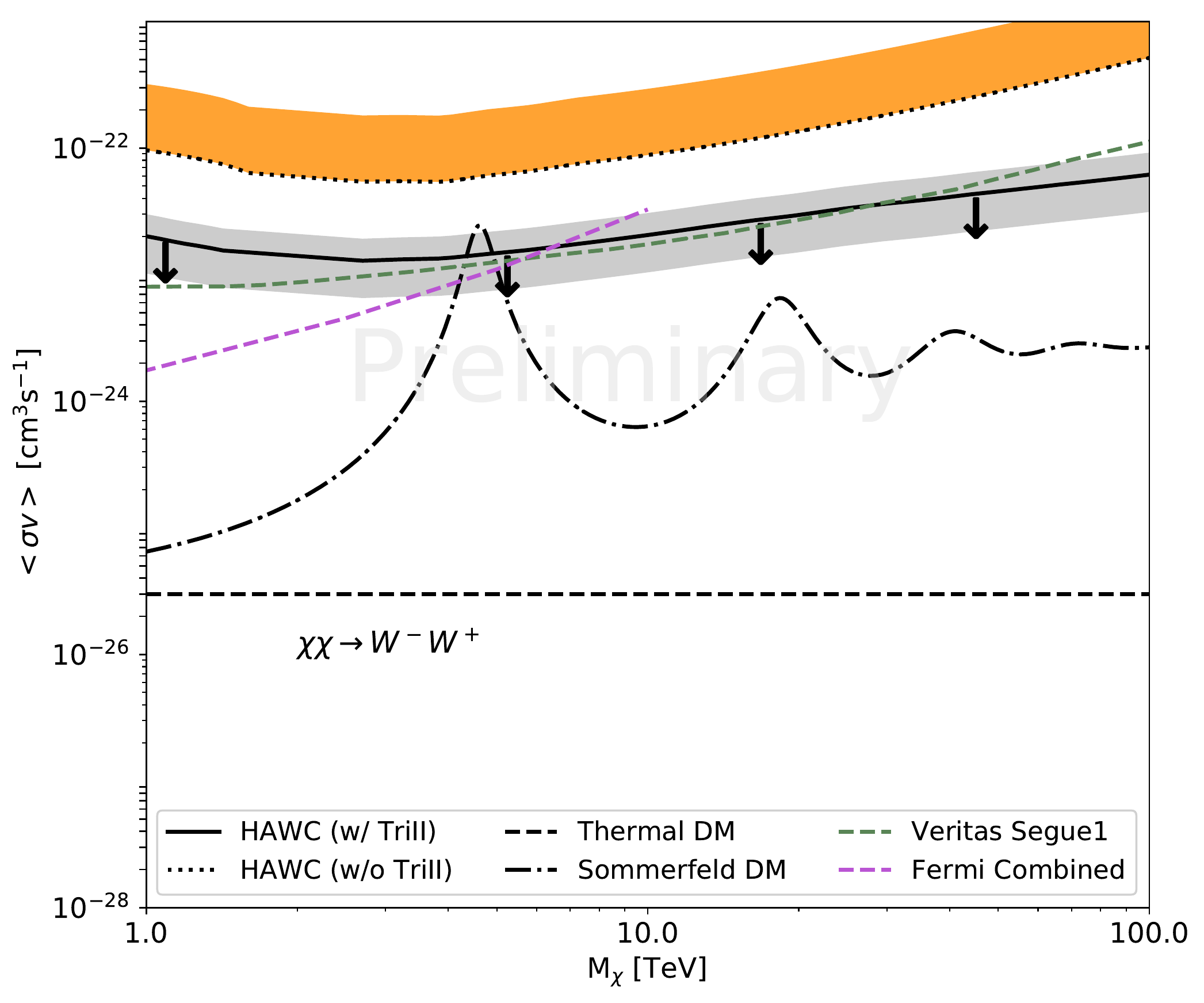}
  \caption{95$\%$ confidence level upper limits on the dark matter annihilation cross-section for the five dark matter annihilation channels considered in this analysis and their comparison of the dark matter annihilation cross-section limits of HAWC to other experimental results for the the $b\bar{b}$, $t\bar{t}$, $\tau^{+}\tau^{-}$, $\mu^{+}\mu^{-}$ and $W^{+}W^{-}$ annihilation channels. The HAWC 507 days limits from data are shown by the black solid line. The dashed black line shows the combined limit using 14 dSphs, excluding Triangulum II. Fermi-LAT combined dSph limits \citep{Fermi2014}, Veritas Segue 1 limits \citep{2017PhRvD..95h2001A}, HESS combined dSph limits \citep{HESS2014} and MAGIC Segue 1 limits \citep{MAGIC2016} are shown for comparison. The same color scheme is used for all the experiment comparison plots.}
\label{fig:overallComparison}
\end{figure*}

The 95$\%$ confidence level upper limits for dark matter annihilating with 100$\%$ branching ratio into the $b \bar b$ channel are shown in the first panel of Figure \ref{15dwarffigs_annihilation}. The individual limits are shown for each dSph considered in this analysis. Figures \ref{15dwarffigs_annihilation} show 15 individual dwarf galaxy limits as well as the combined limit resulting from a joint likelihood analysis. The systematic uncertainties on the observed flux are shown as grey bands on the combined limits. The combined limit is dominated by the influence of three most constraining dSphs with large J-factors with favorable declinations for HAWC: Segue1, ComaB and TriangulumII. The addition of the remaining twelve dSphs does not significantly change the combined annihilation limits. Despite some of them having considerable high J-factors, they are close to the edge of the field of view of HAWC. Thus, HAWC is not sensitive to these sources. The other panels of Figure \ref{15dwarffigs_annihilation} show the same information but for dark matter annihilating with 100$\%$ branching ratio into the $\tau^{+} \tau^{-}$, $\mu^{+}\mu^{-}$, $t\bar t$, and $W^{+}W^{-}$ channels. Comparison of these limits to those of other experiments can be seen in Figure~\ref{fig:overallComparison}. 

In order to directly compare the combined limits on the annihilation cross-section for each individual dark matter channel, the results are shown together in Figure \ref{fig:overallComparison}. The most constraining limit comes from the $\tau^{+}\tau^{-}$ annihilation channel for all dark matter masses considered here.

The combined HAWC limits are compared to limits from four other gamma ray experiments' observations of dSphs, in Figure \ref{fig:overallComparison}. These are the Fermi-LAT combined dSph limits \citep{Fermi2014}, Veritas Segue 1 limits \citep{veritas2012}, HESS combined dSph limits \citep{HESS2014} and MAGIC Segue 1 limits \citep{MAGIC2016}.

For the $b\bar{b}$ channel, Fermi-LAT limit is the most contraining up to $\sim$4 TeV continued with the MAGIC Segue 1 limit up to $\sim$10 TeV. After $\sim$10 TeV, the HAWC combined dSph limit is the most stringent limit for this channel. Similarly, the HAWC combined limits are strongest for the $W^+W^-$ channels for ${M_\chi\gtrsim\mbox{30 {\rm TeV}}}$ and the result is consistent within uncentainties with Veritas Segue 1 limit. For the leptonic $\mu^+\mu^-$ and $\tau^+\tau^-$ channels, the HAWC combined dSph limits are the strongest above a few TeV.

In Figure \ref{fig:overallComparison}.e, dark matter models for thermal relic and Sommerfeld enhanced cross-sections are shown for comparison. For the Sommerfeld enhancement, a weak-scale coupling of 1/35 and a very conservative dark matter velocity of 300\,km/s was assumed. In this work, only $W^+W^-$ annihilation channel is taken into account for the Sommerfeld enhancement since this channel is assured to have dark matter coupled  to gauge bosons \citep{sommerfeld}. At resonances, HAWC limit rules out a dark matter with mass of $\sim$4 TeV, and HAWC limit approaches to corresponding Sommerfeld-enhanced models by 1 order of magnitude for a dark matter with mass of $\sim$20 TeV. Slower dark matter velocity enhances the amplitude of resonances, thus making HAWC results closer to Sommerfeld-enhanced thermal relic.

\begin{longrotatetable}
\begin{deluxetable*}{cccccc}
  \tablecaption{Quasi-model independent 95\% upper flux limits calculated for fifteen dwarf spheroidal galaxies. Spectral index of -2 was assumed for evaluating the flux normalization}
  \label{tbl:upperflux}
  \centering
  \tablewidth{700pt}
  \tabletypesize{\small}
  \tablehead{
  \colhead{Source} & 
  \colhead{0.56-1.78 TeV} &
  \colhead{1.78-5.62 TeV} &
  \colhead{5.62-17.78 TeV} &
  \colhead{17.78-56.23 TeV} &
  \colhead{$>$56.23 TeV TeV}\\
  \colhead{} &
  \colhead{[10$^{-12}$ TeV$^{-1}$cm$^{-2}$s$^{-1}$]} &
  \colhead{[10$^{-14}$ TeV$^{-1}$cm$^{-2}$s$^{-1}$]} &
  \colhead{[10$^{-15}$ TeV$^{-1}$cm$^{-2}$s$^{-1}$]} &
  \colhead{[10$^{-16}$ TeV$^{-1}$cm$^{-2}$s$^{-1}$]} &
  \colhead{[10$^{-17}$ TeV$^{-1}$cm$^{-2}$s$^{-1}$]}
  } 
  \startdata  
    Bootes1          & 3.19 & 5.02  & 2.24 & 1.67 & 1.56 \\
    CanesVenaticiI   & 2.92 & 10.8  & 6.61 & 3.18 & 2.18 \\
    CanesVenaticiII  & 2.47 & 9.18  & 5.38 & 2.25 & 1.45 \\
    ComaB 	     & 1.61 & 6.39  & 1.08 & 0.62 & 0.69 \\
    Draco            & 12.2 & 31.3  & 12.2 & 5.71 & 2.44 \\
    Hercules         & 1.11 & 2.70  & 77.7 & 70.1 & 92.9 \\
    LeoI             & 3.29 & 7.64  & 1.67 & 73.7 & 65.4 \\
    LeoII            & 2.24 & 7.34  & 1.80 & 1.43 & 1.42 \\
    LeoIV            & 4.60 & 10.2  & 4.91 & 1.53 & 88.2 \\
    Segue1           & 1.27 & 7.19  & 5.06 & 2.77 & 2.08 \\
    Sextans          & 5.58 & 15.3  & 5.30 & 2.05 & 1.23 \\
    TriangulumII     & 2.11 & 8.48  & 5.14 & 2.96 & 2.13 \\
    UrsaMajorI       & 9.35 & 15.2  & 4.48 & 1.98 & 1.22 \\
    UrsaMajorII      & 145.0 & 283.4 & 77.8 & 21.2 & 10.2 \\
    UrsaMinor        & 200.5 & 402.6 & 84.1 & 30.4 & 13.8 \\
    \hline
  \enddata
\end{deluxetable*}
\end{longrotatetable}

\subsection{Dark Matter Decay Lifetime Limits}
\begin{figure*}[h]
  \includegraphics[width=0.46\textwidth]{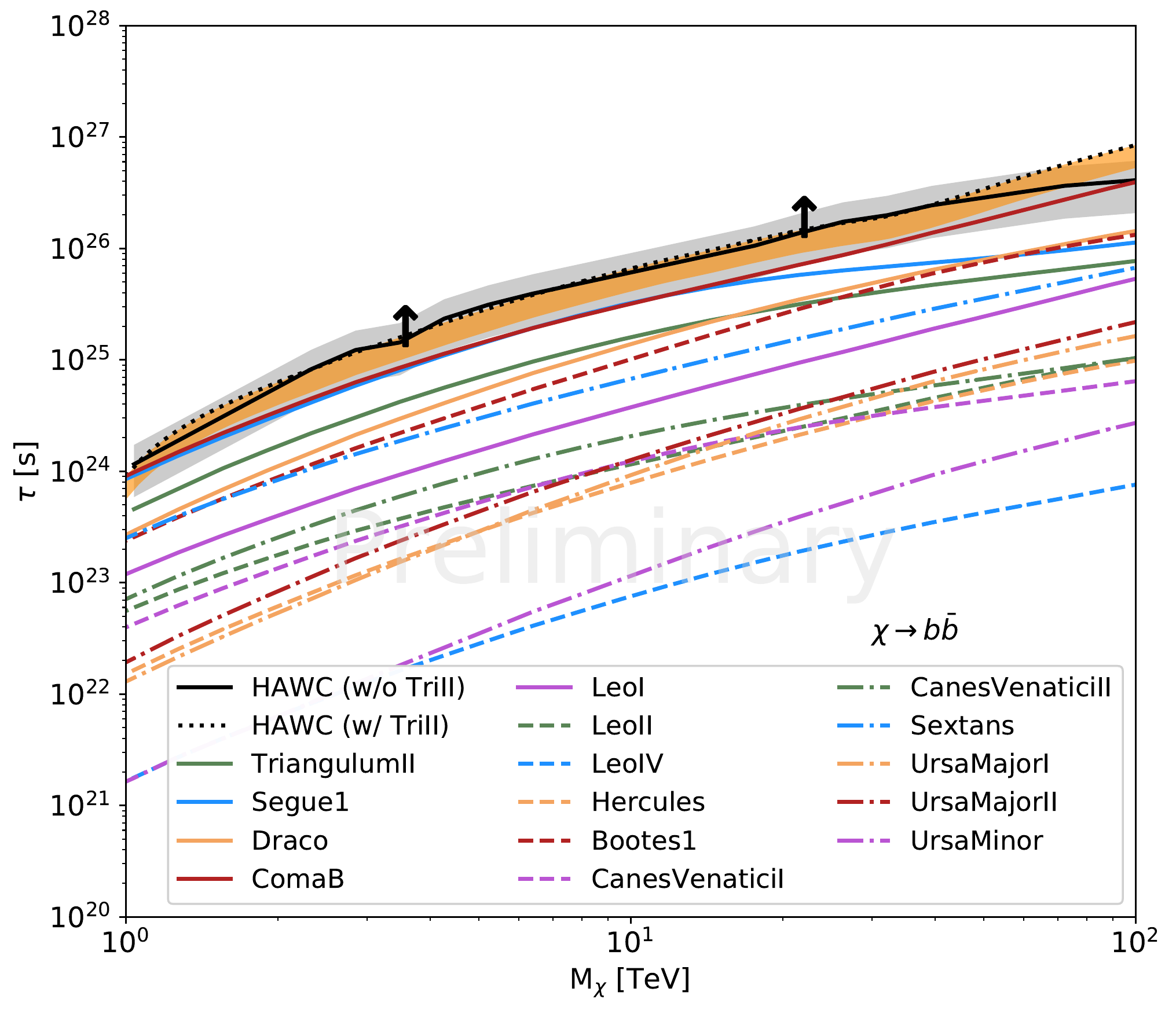}
  \includegraphics[width=0.46\textwidth]{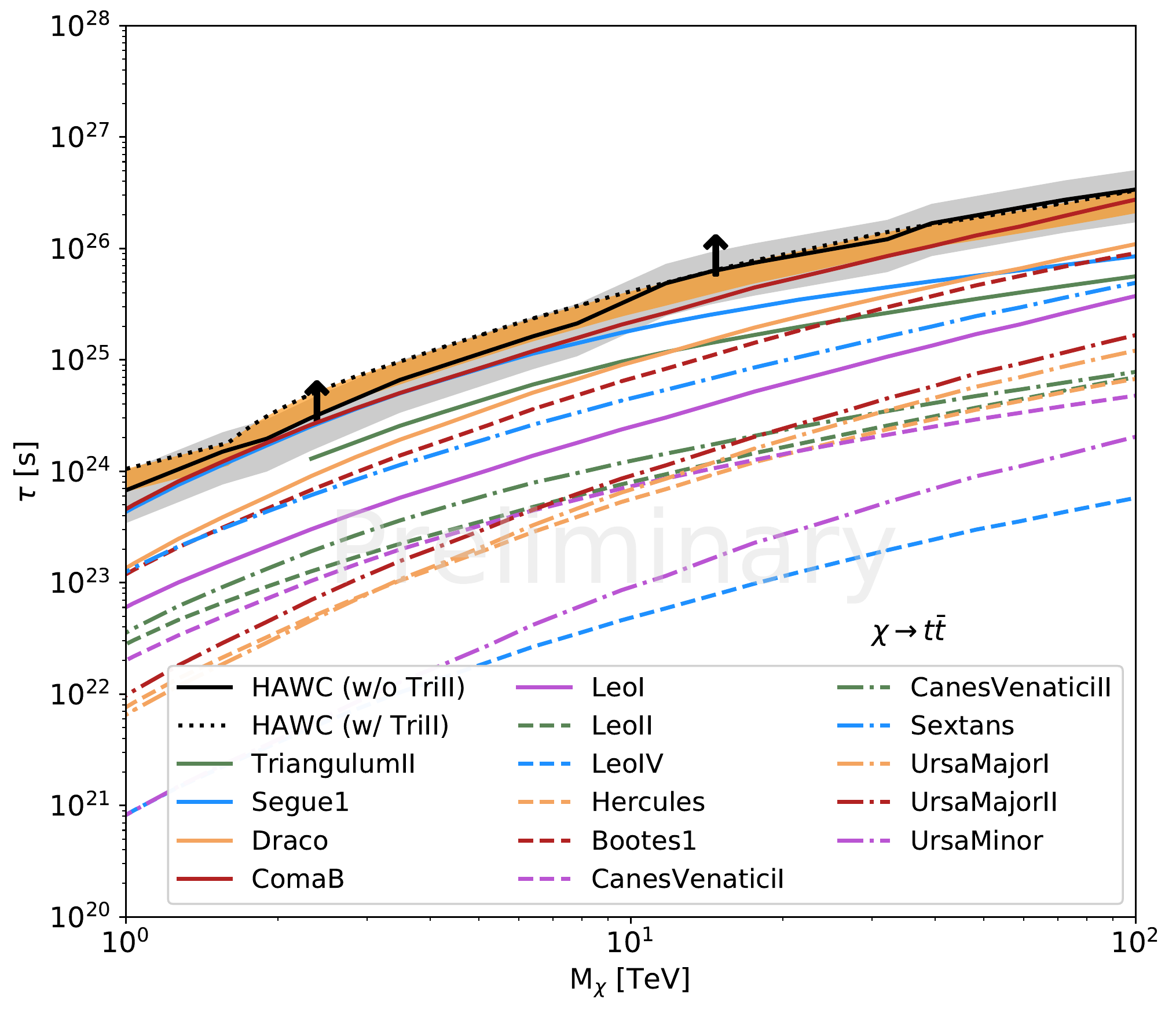}\\
  \includegraphics[width=0.46\textwidth]{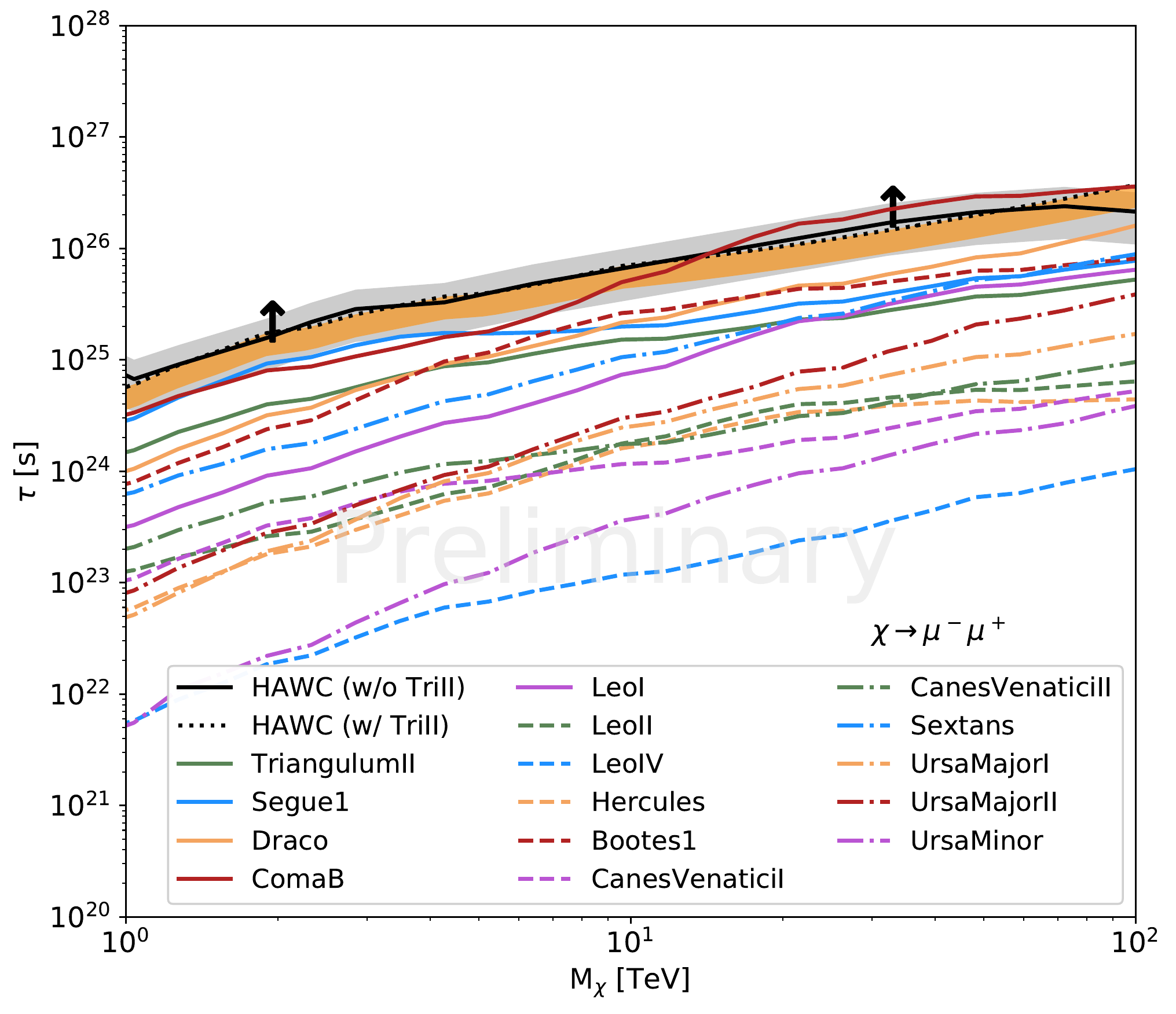}
  \includegraphics[width=0.46\textwidth]{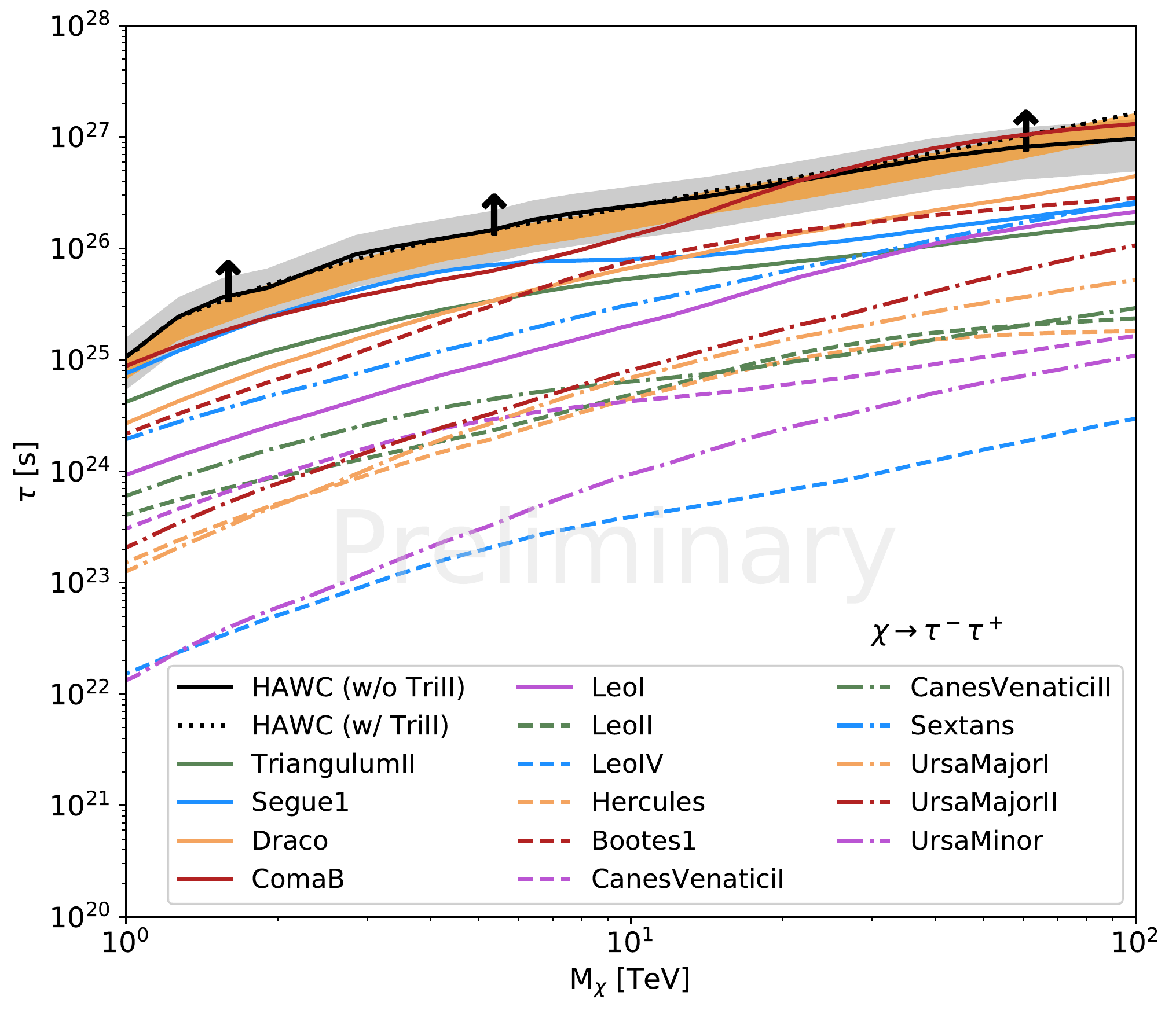}\\
  \includegraphics[width=0.46\textwidth]{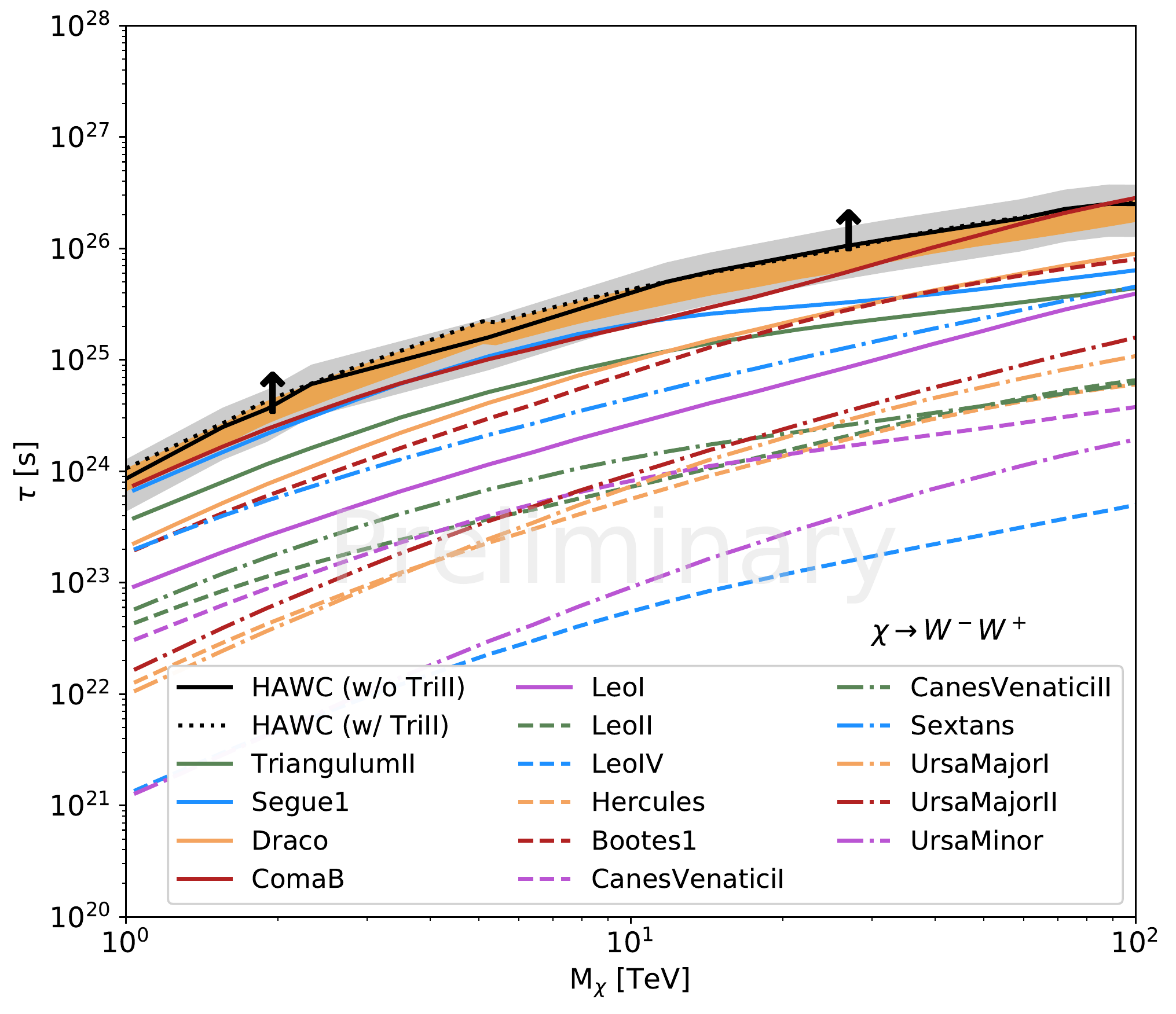}
  \includegraphics[width=0.46\textwidth]{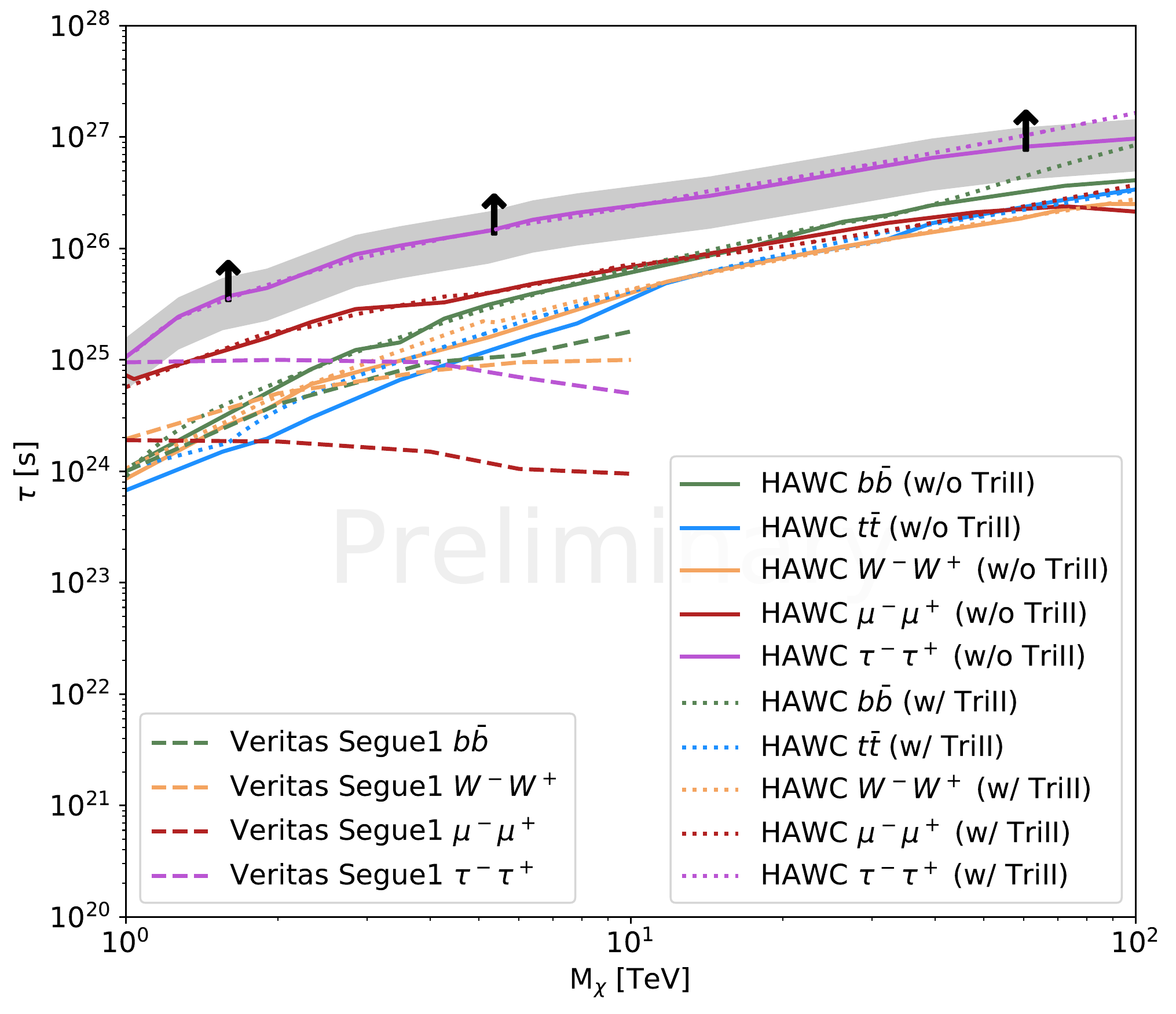}
  \caption{95$\%$ confidence level lower limits on the dark matter decay lifetime for 15 dwarf spheroidal galaxies within the HAWC field of view for the $b \bar b$, $t\bar{t}$, $\tau^{+}\tau^{-}$, $\mu^{+}\mu^{-}$ and $W^{+}W^{-}$ decay channels. The individual limits are shown from the likelihood analysis for all 15 dSphs with the colored dashed and solid lines. The solid black line shows the combined limit using these 15 dSphs resulting from a joint likelihood analysis. The gray band shows the systematic uncertainty on the combined limits. The dashed black line shows the combined limit using 14 dSphs resulting from a joint likelihood analysis, excluding Triangulum II. The gray band shows the systematic uncertainty on the combined limits due to HAWC systematics and dark orange band shows the systematic uncertainty due to D-factor uncertainty. Combined Limits for different channels were compared with Veritas Segue 1 limits \citep{veritas2012} in bottom right panel.
    \label{fig:15dwarffigs_decay}}
\end{figure*}

Dark matter decay lifetime 95\% confidence lower limits calculated for individual channels (with 100\% branching ratio) with 507 days HAWC data are shown in Figure \ref{fig:15dwarffigs_decay}.
The top, middle and bottom panels show the quark channels, lepton channels, and boson channel, respectively, for dark matter masses ranging from 1-100 TeV. Figure \ref{fig:15dwarffigs_decay} shows 15 individual dSph limits, and the combined limit from these 15 dSphs (in black). Similar to the dark matter annihilation results, the limits are driven by Segue 1, Coma Berenices, and Triangulum II, though for decays, Bootes I and Draco also contribute significantly to the combined limits. This is due to the fact that dark matter decay is related to $\int\rho$ (total dark matter mass) compared to $\int\rho^2$ at the source of annihilation or decay. These D-factors of these sources have a different hierarchy of importance than for the J-factors for annihilation.
 
For the lepton channels, due to negative significance at high dark matter masses, the ComaB limits exhibit an increase; however, as in the annihilation limits, the effect is nullified in the joint likelihood analysis. The strongest overall lower limit is attained by the $\tau^+\tau^-$ channel, which is followed by the other lepton channel ($\mu^+\mu^-$). 

\section{Summary and Outlook}

In this analysis presented, we searched for dark matter annihilation and decay signals from fifteen dwarf spheroidal galaxies. We observed no significant excess from these sources. Thus, we calculated individual limits for fifteen dwarf spheroidal galaxies within the HAWC field-of-view using a likelihood ratio analysis method for five dark matter channels. Combined limits from a joint likelihood analysis of all dwarf spheroidal galaxies were also shown. The combined analysis was done to increase the statistical power of the analysis. These are the first limits on the dark matter annihilation cross-section and decay lifetime using data collected from the completed HAWC array. 

The HAWC combined 15 dSph limits were also compared to four other gamma-ray experiments, Fermi-LAT, VERITAS, HESS and MAGIC. While HAWC annihilation cross-section limits with 507 days of data provide complementary results below few TeV, HAWC limits are the most constraining limits above 2-3 TeV and above $\sim$20 TeV for ($b\bar{b}$, $t\bar{t}$, $\mu\mu$, $\tau\tau$) and $W^-W^+$, respectively. As for the decay lifetime limits, HAWC has the only limits above 10 TeV and has the most contraining decay lifetime limits with dSph for all channels.  HAWC decay lifetime limits provide the only limits at dark matter masses higher than $\sim$10 TeV with dSphs. 

We are working on improving our analysis tools for enhancing energy and angular resolution. Moreover, an approved extension of HAWC, consisting of smaller tanks around HAWC array perimeter, is being built. With more data collected, improvements on analysis tools and detector, HAWC is expected to be more sensitive at lower dark matter masses, as well as improve its limits at high masses. 

In addition to the prompt gamma-ray emission discussed here for the calculations of the limits, the charged particles produced in the annihilation or decay may undergo other physical process (such as inverse Compton scattering and Bremsstrahlung) that yields more gamma rays as the charged particles propagate. The gamma-ray flux due to such phenomena peaks at lower energies than the prompt emission. Thus, the gamma-ray flux spectrum will extend down to much lower energies. For the lepton channels, particularly, this effect can increase the dark matter gamma-ray flux significantly. Inclusion of these processes may improve HAWC dark matter limits. The analysis with these additional physics processes will be conducted in the future.

\begin{acknowledgments}
  We acknowledge the support from: the US National Science Foundation (NSF);
  the US Department of Energy Office of High-Energy Physics; the Laboratory
  Directed Research and Development (LDRD) program of Los Alamos National
  Laboratory; Consejo Nacional de Ciencia y Tecnolog\'{\i}a (CONACyT), M{\'e}xico
  (grants 271051, 232656, 260378, 179588, 239762, 254964, 271737, 258865, 243290,
  132197), Laboratorio Nacional HAWC de rayos gamma; L'OREAL Fellowship for Women
  in Science 2014; Red HAWC, M{\'e}xico; DGAPA-UNAM (grants RG100414, IN111315,
  IN111716-3, IA102715, 109916, IA102917); VIEP-BUAP; PIFI 2012, 2013, PROFOCIE
  2014, 2015; the University of Wisconsin Alumni Research Foundation; the
  Institute of Geophysics, Planetary Physics, and Signatures at Los Alamos
  National Laboratory; Polish Science Centre grant DEC-2014/13/B/ST9/945;
  Coordinaci{\'o}n de la Investigaci{\'o}n Cient\'{\i}fica de la Universidad
  Michoacana. Thanks to Luciano D\'{\i}az and Eduardo Murrieta for technical
  support.
\end{acknowledgments}

\clearpage
\appendix
\section{Detailed Calculation of Limits on the Dark Matter Annihilation Cross-section and Decay Lifetime}\label{CalcAppendix}

For each bin of the analysis, we consider the number of observed signal counts $N$ and the number of observed background counts $B$. More detail about the background calculations is explained in~\citep{2017arXiv170101778A}. The number of expected excess counts from a dark matter source $S$ can be calculated by convolving Equation \ref{Flux} with detector energy response at the source position and point-spread function. Then, we define the log-likelihood ratio, test-statistic (TS) as,

\begin{equation} \label{likely}
  TS = -2\ln\left(\frac{\mathcal{L}_0}{\mathcal{L^{\rm max}}}\right)
\end{equation}
where $\mathcal{L}_0$ is the null hypothesis (no DM model) likelihood and $\mathcal{L^{\rm max}}$ is the alternative hypothesis (with DM model) likelihood, evaluated at the value of the cross-section which maximizes the likelihood. Both likelihoods are taken to be Poisson distributions in each bin:
\begin{equation}
  \mathcal{L} = \prod_{i}\frac{(B_i+S_i)^{N_i} \exp\left[{-(B_i+S_i)}\right]}{N_i!}
\end{equation}
where $S_i$ is the sum of expected number of signal counts corresponding to a annihilation cross-section or a decay lifetime, $B_i$ is the number of background counts observed, and $N_i$ is the total number of counts observed. Since negative dark matter cross-sections and lifetimes are physically not allowed, the value of $S_i$ is restricted to positive values. Therefore, for sources which are within underfluctuations of the background, the value of $S_i$ which maximizes the likelihood is $S_i=0$, consistent with no gamma rays from dark matter annihilation or decay. For these underfluctuations, we find
\begin{equation}
  \mathcal{L^{\rm max}} = \mathcal{L}_0 =  
    \prod_{i}\frac{B_i^{N_i} \exp\left[{-B_i}\right]}{N_i!}\enspace.
\end{equation}
This gives a $TS$ value of zero for these underfluctuations.


The likelihood is calculated over all spatial bins near the source and all HAWC analysis bins (fHit bins) \citep{2017arXiv170101778A}.
The spatial binning we use in this analysis spans 0.0573 degrees corresponding to 9.986$\times$10$^{-7}$ steradian, smaller than the point-spread function of the detector. Also, because the dark matter profile peaks strongly toward the center of each source, as discussed in section~\ref{DMdenssec}, much faster than the HAWC point-spread function, we expect negligible difference between a point-source analysis and one treating the dwarfs as extended sources.

\subsection{95$\%$ Confidence Level Limit Calculation}

Although the null hypothesis is a good approximation for our sources, we have set up our likelihood calculation to be robust to possible statistical fluctuations or small positive indications of sources. To do so, we introduce the parameter $TS_{max}$, which is the maximum value of the $TS$ for a given dark matter mass and a given annihilation or decay channel. TS$_{max}$ corresponds to an annihilation cross-section, $\langle \sigma_{A} v \rangle_{max}$ or decay lifetime, $\tau_{max}$; however the subscripts for cross-section and lifetime should not be interpretted as their maximum values. In the case of the dwarf spheroidal galaxies, $TS_{max}$ is zero or very close to zero. 

We calculate an upper limit on the annihilation cross-section or a lower limit on the decay lifetime by setting $95\%$ confidence level (CL) limit.
For this confidence level, we define the parameter, $TS_{95}$,
\begin{equation} \label{TS95}
TS_{95} = \sum_{bins}\left[2 N \ln \left ( 1 + \frac{\xi S_{ref}}{B} \right) - 2 \xi S_{ref} \right]
\end{equation}
where we scale the number of expected signal counts from a source by a scale factor $\xi$ and $S_{ref}$ is the expected number of excess counts in a bin due to a dark matter source with a reference annihilation cross-section, $\langle \sigma_{A} v \rangle_{ref}$ or decay lifetime, $\tau_{ref}$. This allows us to calculate the decreasing likelihood of observing higher numbers of gamma rays being emitted from a potential dark matter source. We find $\xi$ such that
\begin{equation} \label{DeltaTS}
2.71 = \Delta TS = TS_{max} - TS_{95}
\end{equation}
where $\Delta TS$ is the difference between $TS_{max}$ and $TS_{95}$. For a one-sided 95\% CL limit, $\Delta TS = 2.71$ corresponds to a likelihood which can be excluded at 95\% CL~(\cite{Fermi2014}).

Once the scale parameter $\xi$ is found, we then scale the reference annihilation cross-section ($\langle \sigma_{A} v \rangle_{ref}$) or decay lifetime ($\tau_{ref}$) that was used to calculate the dark matter gamma-ray flux, for a given $M_{\chi}$ and annihilation channel, by the same parameter $\xi$. Thus our 95\% CL limit on the annihilation cross-section becomes:

\begin{equation} \label{limit}
\langle\sigma_{A} v \rangle_{95\%} = \xi \times \langle \sigma_{A} v \rangle_{ref}
\end{equation}

and the limit on the decay lifetime is

\begin{equation} \label{limit}
\tau_{95\%} = \tau_{ref} / \xi \enspace.
\end{equation}

\subsection{Joint Likelihood Analysis}

For the joint likelihood analysis of many dwarf spheroidal galaxies, the same likelihood analysis procedure is followed as described in the above section. However, now Equation \ref{DeltaTS} becomes

\begin{equation} \label{TScombined}
2.71 = \Delta TS = TS_{max}- \sum_{bins}\left[2 N^{tot} \ln \left ( 1 + \frac{\xi^{tot} S_{ref}^{tot}}{B^{tot}} \right) - 2 \xi^{tot} S_{ref}^{tot} \right]
\end{equation}
where $N^{tot}$ is the total events in each bin from data summed over all the dSphs, $B^{tot}$ is the summed total of background counts from each dSph and $S_{ref}^{tot}$ is the total expected number of counts in each bin for the reference annihilation cross section or decay lifetime for all the dSphs. The same procedure is then followed: we find $\xi^{tot}$ by imposing the condition in Equation \ref{TScombined}, such that the difference between $TS_{max}$ and $TS_{95}$ is equal to 2.71 for the combined analysis. Once $\xi^{tot}$ is found, we can then scale $\langle \sigma_{A} v \rangle_{ref}$ or $\tau_{ref}$ in order to set our constraint on the combined analysis of the dSphs:

\begin{eqnarray}
\langle \sigma_{A}v\rangle_{95\%, Combined} &=& \xi^{tot} \times \langle \sigma_{A}v\rangle_{ref}\\
\tau_{95\%, Combined} &=& \tau_{ref} / \xi^{tot}\enspace.
\end{eqnarray}

\section{Calculating Model Limits from Tabulated Flux Limits}\label{LimitAppendix}

Although the interpretation of the limits in this paper is primarily dark matter focused, it cannot be stressed enough that first and foremost, these are flux limits. As discussed in section~\ref{flimits} and shown in table~\ref{tbl:upperflux}, we have calculated generic flux limits which can be used to constrain dark matter models. Here, we discuss how to use these flux limits to constrain dark matter models not considered in this paper.

The limits in table~\ref{tbl:upperflux} are 95\% CL limits in each energy range for each source. Therefore, they correspond to $\Delta TS = 2.71$, as discussed in appendix~\ref{CalcAppendix}. Because the sources have low statistical significance, this can be approximated as $TS_{95} = -2.71$. In the Gaussian statistical regime, which is valid for large number of counts, the test statistic $TS$ is proportional to the square of the signal flux. These properties can be used to get an approximate limit for any model-specific flux spectrum from the data in table~\ref{tbl:upperflux}.

For each limiting energy bin $i$, let $F^{\rm lim}_i$ be the limiting flux in that bin. Select a reference value of $\langle\sigma_{A}v\rangle^{\rm ref}$ (for annihilation) or $\tau^{\rm ref}$ (for decay). The limit will not depend on this choice of reference cross-section or lifetime, but it will be necessary for calculating flux in this calculation. For the flux spectrum to be constrained ($F^{\rm model}$), define
\begin{eqnarray}
F^{\rm model,ref}_i &\equiv& F^{\rm model}(E_i;\langle\sigma_{A}v\rangle^{\rm ref})\\
{\rm or}\enspace F^{\rm model,ref}_i &\equiv& F^{\rm model}(E_i;\tau^{\rm ref})
\end{eqnarray}
depending on whether the calculation is for annihilation or decay. Here $E_i$ is the energy of the limit bin $i$. The calculated limit will be done as in appendix~\ref{CalcAppendix} as
\begin{eqnarray}
\langle \sigma_{A}v\rangle_{95\%} &=& \xi \times \langle \sigma_{A}v\rangle_{ref}\label{limitxiann}\\
\tau_{95\%} &=& \tau_{ref} / \xi\enspace\label{limitxidec}.
\end{eqnarray}

For a given value of $\xi$, the approximate test statistic for a single energy bin is then
\begin{equation}
2.71\left(\frac{\xi F^{\rm model,ref}_i}{F^{\rm lim}_i}\right)^2\approx TS\enspace.
\end{equation}
These single-bin $TS$ values can be summed over all energies to get a total constraint on the spectrum over all energies.
\begin{equation}
\sum_{i}2.71\left(\frac{\xi F^{\rm model,ref}_i}{F^{\rm lim}_i}\right)^2\approx TS^{\rm tot}\enspace.
\end{equation}
For a 95\% CL limit on the spectrum, we need to find the value of $\xi$ which gives $TS^{\rm tot}=2.71$. This value is
\begin{equation}
\xi=\left[\sum_{i}\left(\frac{F^{\rm model,ref}_i}{F^{\rm lim}_i}\right)^2\right]^{-1/2}
\end{equation}
and the corresponding approximate 95\% CL limits on the cross-section or lifetime are then calculated from equation~\ref{limitxiann} or~\ref{limitxidec}.
To calculate the approximate combined limit from all dwarf galaxies, simply sum $i$ over all energy bins of all sources, making sure to calculate $F^{\rm model,ref}_i$ for each source based on its J- or D-factor. 

These approximate limits have been checked over a variety of spectra and agree with the full HAWC analysis calculation to better than 50\%. It should be noted that the HAWC sensitivity depends on the spectrum of the gamma-ray source being studied, and the HAWC energy resolution is very broad. Additionally, although the statistical significance of these sources is roughly Gaussian and the statistical fluctuations are small, some calculation error is still introduced from these effects. Therefore, limits calculated in this way are only approximate, with errors in the tens of percent. A more correct analysis of a candidate dark matter spectrum requires a full calculation through the HAWC analysis chain.

\clearpage
\bibliography{bibliography}

\end{document}